\documentclass[10pt,preprint]{aastex}
\def\insertplot#1#2#3#4#5#6#7{
\vskip 10pt\nobreak\hbox to \hsize{\hss\dimen0=#3in\hbox to #6\dimen0{%
\dimen0=#2in\vbox to #6\dimen0{\vss
\special{ps: plotfile #1}
\special{ps::[end]
  PGPLOT restore
}
}\hss}\hss}\vskip 10pt}
\usepackage{graphicx}
\usepackage{ulem}
\usepackage{rotating}
\usepackage{lscape}
\begin{document}

\title{The Formation and Evolution of Planetary Systems: 
Placing Our Solar System in Context with Spitzer}
\author{Michael R. Meyer} \affil{Steward Observatory, The University of Arizona}
\authoremail{mmeyer@as.arizona.edu}
\author{Lynne A. Hillenbrand} \affil{California Institute of Technology}
\author{Dana Backman} \affil{SOFIA/SETI Institute}
\author{Steve Beckwith} \affil{Space Telescope Science Institute} \affil{Department of Physics and Astronomy, Johns Hopkins University}
\author{Jeroen Bouwman} \affil{Max--Planck--Institut f\"ur Astronomie, Heidelberg}
\author{Tim Brooke} \affil{California Institute of Technology}
\author{John Carpenter} \affil{California Institute of Technology} 
\author{Martin Cohen} \affil{Radio Astronomy Laboratory, University of California--Berkeley} 
\author{Stephanie Cortes} \affil{Steward Observatory, The University of Arizona}
\author{Nathan Crockett} \affil{National Optical Astronomy Observatories}
\author{Uma Gorti} \affil{NASA-Ames Research Center, Theory Branch} 
\author{Thomas Henning} \affil{Max--Planck--Institut f\"ur Astronomie, Heidelberg}
\author{Dean Hines} \affil{Space Science Institute} 
\author{David Hollenbach} \affil{NASA-Ames Research Center, Theory Branch} 
\author{Jinyoung Serena Kim} \affil{Steward Observatory, The University of Arizona} 
\author{Jonathan Lunine} \affil{Lunar and Planetary Laboratory, The University of Arizona} 
\author{Renu Malhotra} \affil{Lunar and Planetary Laboratory, The University of Arizona}
\author{Eric Mamajek} \affil{Harvard--Smithsonian Center for Astrophysics} 
\author{Stanimir Metchev} \affil{The University of California, Los Angeles}
\author{Amaya Moro-Martin} \affil{Department of Astrophysical Science, Princeton University} 
\author{Pat Morris} \affil{Herschel Science Center, IPAC} 
\author{Joan Najita} \affil{National Optical Astronomy Observatories} 
\author{Deborah Padgett} \affil{Spitzer Science Center} 
\author{Ilaria Pascucci} \affil{Steward Observatory, The University of Arizona}
\author{Jens Rodmann} \affil{Max--Planck--Institut f\"ur Astronomie, Heidelberg} 
\author{Wayne Schlingman} \affil{Steward Observatory, The University of Arizona}
\author{Murray Silverstone} \affil{Steward Observatory, The University of Arizona}
\author{David Soderblom} \affil{Space Telescope Science Institute} 
\author{John Stauffer} \affil{Spitzer Science Center}
\author{Elizabeth Stobie} \affil{Steward Observatory, The University of Arizona}
\author{Steve Strom} \affil{National Optical Astronomy Observatories}
\author{Dan Watson} \affil{Department of Physics and Astronomy, The University of Rochester} 
\author{Stuart Weidenschilling} \affil{Planetary Science Institute} 
\author{Sebastian Wolf} \affil{Max--Planck--Institut f\"ur Astronomie, Heidelberg}
\author{Erick Young} \affil{Steward Observatory, The University of Arizona}

\received{}
\revised{}
\accepted{}
\journalid{}{}
\articleid{}{}

\begin{abstract}

We provide an overview of the Spitzer Legacy Program 
``Formation and Evolution of Planetary Systems" (FEPS) which was proposed 
in 2000, begun in 2001, and executed aboard the Spitzer Space Telescope
between 2003 and 2006. 
This program exploits the sensitivity of Spitzer to carry out mid-infrared 
spectrophotometric observations of solar-type stars.  
With a sample of $\sim$328 stars ranging in age from 
$\sim$3 Myr to $\sim$3 Gyr, we trace the evolution of circumstellar gas 
and dust from primordial planet-building stages in young circumstellar disks
through to older collisionally generated debris disks.  When completed,
our program will help 
define the time scales over which terrestrial and gas giant planets are built,
constrain the frequency of planetesimal collisions as a function of time, 
and establish the diversity of mature planetary architectures. 

In addition to the observational program,
we have coordinated a concomitant theoretical effort aimed at understanding
the dynamics of circumstellar dust with and without the effects of embedded 
planets, dust spectral energy distributions, and atomic and molecular gas line 
emission.  Together with the observations, these efforts will provide 
astronomical context for understanding whether our Solar System -- and its 
habitable planet -- is a common or a rare circumstance.  
Additional information about the FEPS project can be found 
on the team website: \begin{verbatim}http://feps.as.arizona.edu/ \end{verbatim}

\end{abstract}


\section{Introduction}

The Spitzer Space Telescope (Werner et al. 2004), formerly SIRTF the
{\bf S}pace {\bf I}nfra{\bf R}ed {\bf T}elescope {\bf F}acility,
is an 85 cm cryogenic space observatory in earth--trailing orbit. The 
observatory was launched in August of 2003 and has an estimated mission 
lifetime of 5$+$ years. There are three science instruments on--board: 
IRAC (Fazio et al. 2004), IRS (Houck et al. 2004), and MIPS (Rieke et al. 2004)
which together provide the capability for imaging and spectroscopy
from 3.6--160 $\mu$m.  The Legacy Science Program was established before 
launch with two goals: to enable large scale programs of
broad scientific and public interest, and to provide access to uniform 
and coherent datasets 
as rapidly as possible in support of General Observer (GO) proposals,
given the limited lifetime of the mission. 
The Formation and Evolution of Planetary Systems (FEPS) Spitzer Legacy Science 
Program is one of six such original programs 
(for descriptions of the others see 
Evans et al. 2003; Benjamin et al. 2003; Kennicutt et al. 2003; 
Londsdale et al. 2003; Dickenson et al. 2003) and uses 350 hours of Spitzer
observing time.  FEPS builds upon the rich heritage of 
Spitzer's ancestors in space: the international 
all-sky mid-infrared survey telescope
IRAS (the Infrared Astronomical Satellite, 1983--1985) and ESA's 
pointed mission ISO (the Infrared Space Observatory, 1995--1999), 
and complements Guaranteed Time Observer (GTO) and GO 
programs also being pursued with Spitzer. 
 
In a single sentence, FEPS is a comprehensive study of the evolution 
of gas and dust in the circumstellar environment.  The scientific motivation for FEPS 
lies in the fragmented but compelling evidence for dusty circumstellar 
material surrounding stars spanning a wide range of ages, from young 
pre-main sequence stars to those as old as, and even older than, the Sun.

At young ages, incontrovertible evidence assembled over the past three decades 
(based on data from ultraviolet through millimeter wavelengths),
suggests that most stars are surrounded at birth by accretion disks that are
remnants of the star formation process itself (e.g. Beckwith and Sargent, 1996).
The revelations provided by IRAS, and later ISO, led to a nearly complete 
census of optically--thick disks 
within 100-200 pc, and in the case of ISO revealed their rich dust mineralogy 
and gas content (see Lorenzetti 2005 and Molster \& Kemper 2005 for reviews). 
That at least some of these disks build planets
has become clear from radial velocity and photometric studies
revealing M~sin$i$=0.02-15 M$_J$ planets orbiting well over one hundred
nearby stars (e.g. Marcy et al. 2005).  
At older ages, IRAS and ISO revealed the presence around dozens of 
main sequence stars of micron-sized grains.  These dusty ``debris" disks are 
produced in collisions between asteroid-like bodies with orbits that are 
dynamically stirred by planets (e.g. Lagrange et al. 2000).  
Subsequently, several of these
disks were spatially resolved at optical, infrared, and millimeter 
wavelengths (e.g. Kalas, Liu, \& Mathews 2004; Weinberger et al. 1999; 
Greaves et al. 1998) revealing structure consistent with the 
planetary perturber interpretation.  

The connections between planets,
debris disks, and the dusty and gaseous disks near-ubiquitously found in 
association with recently formed young stars are tantalizing,
but not yet unequivocally established.
Understanding the evolution of young circumstellar dust
and gas disks as they transition through the planet--building phase
requires the hundred-fold enhancement in sensitivity and increased
photometric accuracy offered by Spitzer at mid- and far-infrared wavlengths.
For main sequence stars, while IRAS discovered 
the prototypical debris disks (see Backman \& Paresce 1993 for a review) 
and ISO made additional surveys (see de Muizon 2005 for a review), 
neither IRAS nor ISO was sensitive enough to detect dust in solar 
systems older than a few hundred Myr for any but the nearest tens of stars
\footnote{The IRAS and ISO observatories were able to study representative
samples of A-type stars, but not G-type stars.}.
Spitzer, by contrast can detect orders of magnitude smaller dust masses: 
for a solar-type star at 30 pc, down to  
$\sim 10^{20}$ kg or $\sim 10^{-5}$ M$_{Earth}$ in micron to sub-millimeter
size grains at 50 K, only an order of magnitude above the dust mass
inferred for our own present-day Kuiper Belt,
and $\sim 10^{17}$ kg or $\sim 10^{-8}$ M$_{Earth}$ at 150 K,
only an order of magnitude above the dust mass in our
present-day asteroid  belt plus zodiacal cloud.

The FEPS program is designed to study circumstellar dust properties
around a representative sample of solar-type stars.  Included are 328 stars
chosen to probe the suspected direct link between disks commonly found 
around pre--main sequence stars $<$ 3.0 Myr old and our 4.56 Gyr old
Sun and Solar System.
Specifically, we trace the evolution of circumstellar material at ages 
3-10 Myr when stellar accretion from the disk terminates,
to 10-100 Myr when planets achieve their final masses via coalescence 
of solids and accretion of remnant molecular gas, to 100-1000 Myr when 
the final architecture of solar systems takes form and frequent collisions 
between remnant planetesimals produce copious quantities of dust, and finally, 
to 1-3 Gyr mature systems in which planet-driven activity of planetesimals 
continues to generate detectable dust. 
Our sample is distributed uniformly in log-age from 3 Myr to 3 Gyr.
We probe the full range of dust disk optical--depths diagnostic of
the major phases of planet system formation and evolution, including
primordial disks (those dominated by ISM grains 
in the process of agglomerating into planetesimals) and debris disks 
(those dominated by collisionally generated dust) like our own.

Our strategy is to obtain for all 328 stars in our sample carefully 
calibrated spectral energy distributions (SEDs) using all three
Spitzer intstruments.  A high-resolution spectroscopic survey limited 
to the younger targets establishes the gas content.  In addition to
insight into problems of fundamental scientific and philosophical interest,
the FEPS Legacy Science Program provides a rich database for follow-up 
observations with Spitzer, with existing and future ground-based facilities, 
as well as SIM and JWST, and eventually TPF.  FEPS complements and 
motivates many existing Spitzer GTO and GO programs.

\section{Science Strategy}

We take advantage of Spitzer's unprecedented mid-infrared sensitivity and 
hence its unique ability to detect the photospheres of solar-type stars 
out to distances of several tens of pc.  Spitzer observations of excesses 
above those photospheres are indicative of dust located at a range 
of orbital separations, from analogs of
the terrestrial zone, to the gas-giant zone, to the Kuiper belt zone 
in our own Solar System.  Such observations are important 
in the search for exo-solar planetary systems -- either those 
in formation from primordial dust and gas disks, or those which later perturb 
planetesimals into crossing orbits that collisionally cascade to produce 
dusty debris disks.  For a given system, the mass in small grains to which
Spitzer is sensitive is first expected to decrease with time as planet formation
begins, then increase on a relatively rapid time scale (few Myr) as the 
debris phase begins, and finally decrease on a much longer time scale (many Gyr)
as the disk slowly grinds itself down and grains are removed via 
radiative and mechanical effects.

With the goal of understanding how common or rare
the evolutionary path taken by our Solar System might have been,
we have initiated a Spitzer survey of F--G--K (solar-type) stars.
First, we study the formation of planetary embryos in a 
survey of post--accretion circumstellar dust disks.  We aim to
understand the evolution of disk properties (mass and radial 
structure) and dust properties (size and composition) 
during the main phase of planet--building and early solar system 
evolution from 3--100 Myr.  Second, we study
the growth of gas giants in a sensitive search for warm molecular gas 
at the level of $>$2 $\times$ 10$^{-4}$ M$_{\odot}[H_2]$ at 70--200 K, 
in a sub--sample of the targets from our 
dust disk survey.  Our goal is to constrain directly the time available
for embryonic planets to accrete gas envelopes.
Third, we investigate mature solar system evolution by
tracing 100 Myr to 3 Gyr old dust disks generated through collisions 
of planetesimals.  Through our analysis we hope to infer the locations 
and masses of giant planets $>0.05$ M$_{\rm Jupiter} \approx $ 1 M$_{\rm Uranus}$ 
through their action on the remnant disk.

Our large sample enables us to measure the 
{\it mean properties} of evolving dust disks and discover the
{\it dispersion} in evolutionary time scales.  Further, we can search for
relations between inferred dust evolutionary path and stellar properties
such as metallicity and multiplicity.   In the following subsections we
detail our science strategy.

\subsection{Formation of Planetary Embryos}

Our experiment begins as the disks are
making the transition from optically thick to thin, the point at which
all of the disk's mass first becomes detectable through 
observation.  The FEPS goals are to:

\begin{list}{}{\itemsep=-0.05truein  \leftmargin 12pt}
\item{$\bullet$}
constrain the initial structure and composition of
post-accretion, optically thin disks;
\item{$\bullet$}
trace the evolution of disk structure, composition, and mass over time;
\item{$\bullet$}
characterize the time scales over which primordial disks dissipate
and debris disks arise;
\item{$\bullet$}
measure changes in the dust particle size distribution due to 
coagulation of interstellar grains at early stages and shattering associated 
with high-speed planetesimal collisions at later early-debris stages;
\item{$\bullet$}
infer the presence of newly formed planets at orbital radii of 0.3-30 AU.
\end{list}

Photometric observations from 3.6-160$\mu$m with Spitzer probe 
temperatures (radii) encompassing the entire system of planets 
in our Solar System.  In Figure 1 we show the mass sensitivity of 
Spitzer as a function of wavelength, indicating the mass in 
small grains that Spitzer can detect as a function of orbital radius. 
Detailed spectrophotometry in the range 5.3--40 $\mu$m 
permits a search for gaps in disks caused by the dynamical
interaction of young gas giant planets and the particulate disk
from 0.2--10 AU.  This extends from just outside the innermost
radii of the exo-solar ``hot Jupiters'' (thought to have suffered significant
orbital migration in a viscous accretion disk) to the gas giants 
of our Solar System (thought to have formed beyond the ``ice line''). 

\begin{figure}
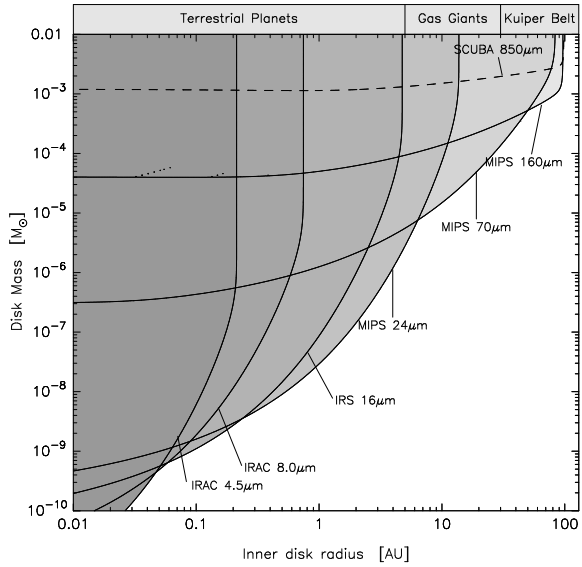

\insertplot{mmeyer_etal_feps_fig1.ps}{4.0}{4.0}{0.30}{3.0}{0.4}{1}
\caption{
Spitzer sensitivity to mass
in small grains as a function of radius in a hypothetical circumstellar disk
surrounding a sun--like star at a distance of 30 pc for integration
times typical for FEPS.
For emission from small grains in radiative equilibrium,
each radius in the disk corresponds to a specific dust temperature.
From simple blackbody considerations, shorter Spitzer wavelengths
probe warmer dust (at smaller orbital radii) while longer Spitzer wavelengths
probe cooler dust (at larger orbital radii), as indicated.
}
\end{figure}

Mid--infrared spectroscopic observations are sensitive to dust 
properties including size distribution and composition. They thus
probe physical conditions in the disk.  
From observations in the 5.3--40 $\mu$m spectral region
we determine the relative importance of broad 
features attributed to amorphous silicates 
(ubiquitous in the ISM) compared to numerous narrow
features due to crystalline 
dust (observed only in circumstellar environments).  In this way, 
we can look for evidence of, e.g. radial mixing in the disk 
since the temperature required to anneal grains ($>$ 1000K) 
is substantial higher than that
inferred for the continuum emitting material ($\sim$ 300K).  
Further, the shape and strength of specific mid-infrared 
spectroscopic features provide constraints on the fractional 
contribution of each grain population to the total opacity, 
necessary for estimating dust mass surface densities
(see section 4.2). 

\subsection{Growth of Gas Giants}

Next, we have undertaken the most sensitive survey to date 
of atomic and molecular gas in post--accretion disk
systems.  In order to characterize gas dissipation and
to place limits on the time available for giant planet formation
we obtain high spectral resolution (R=600) data from 10--37 $\mu$m 
with the IRS of 35 stars selected from our dust disk survey sample.
The data include the S(0) 28.2 $\mu$m, S(1) 17.0 $\mu$m, 
and S(2) 12.3 $\mu$m H$_2$ lines as
well as strong atomic lines such as [SI] 25.23 $\mu$m, [Si II] 34.8 $\mu$m, 
and [FeII] 26 $\mu$m (Gorti \& Hollenbach 2004; 
Hollenbach et al. 2005).
We focus on the post-accretion epochs from 3--100 Myr 
to examine whether gas disks persist after
disk accretion onto the star has ceased and planetesimal agglomeration
has removed the dust disk, potentially
providing ``nucleation sites'' for gas giant planet formation.

Understanding gas--dust dynamics is crucial to our ability to 
derive the time scales important in planet formation and evolution. 
The dust and gas experiments being conducted at young ages ($<$100 Myr)
have an important synergy in furthering this understanding because 
dust dynamics are controlled by gas drag rather than
radiation pressure when the gas-to-dust mass ratio is $>$0.1,
while it is the presence of dust that mediates gas heating 
and therefore detectability. 
If the gas to dust ratio is low (the dust opacity per gas particle high) and 
the gas and dust are at similar temperatures, the detection of gas lines by 
Spitzer becomes difficult due to the small ratio of line to continuum.  
However, the theoretical models of Gorti \& Hollenbach (2004) show 
that in many instances the dust opacity is sufficiently low and the 
gas temperature sufficiently high ($> 100$ K) that small 
quantities ($< 0.1$ M$_J$) of gas, if present,  can be detected by 
Spitzer around nearby disks with optically thin dust (see 
section 4.4). 

\subsection{Mature Solar System Evolution}

Finally, we conduct a study of second generation ``debris disks''.
The presence of {\it any} observable circumstellar dust around 
stars older than the maximum lifetime of a primordial dust 
disk (the sum of the to--be--determined gas dissipation time scale 
and the characteristic Poynting--Robertson drag time scale)
provides compelling evidence not only for large reservoirs of planetesimals
colliding to produce the dust, but also for the existence of massive planetary
bodies that dynamically perturb planetesimal orbits inducing frequent 
collisions.  

We have undertaken the first comprehensive survey of F5--K5
stars with ages 100 Myr to 3 Gyr that is sensitive to dust disks comparable
to those characteristic of our own Solar System throughout
its evolution.  We chart the history of our Solar System from
100--300 Myr, the last phase of terrestrial and ice giant
((Uranus- and Neptune-like) planet--building, through
0.3--1 Gyr, bracketing the ``late heavy bombardment'' impact peak
which might have had an effect on the early 
evolution of life on Earth, and finally over
1.0--3.0 Gyr, examining the diversity of evolutionary paths among
mature planetary system. 
Spectroscopic observations in the range 5.3--40 $\mu$m enable diagnosis of gaps
caused by giant planets 
and estimates of dust size and composition which
translate directly into constraints on the mass opacity coeffients 
for the dust (Miyake \& Nakagawa, 1993) 
as well as Poynting-Robertson drag time scales (Backman e\& Parsece, 1993). 

\section{Survey Preparation and Execution}

\subsection{Observing Strategy}
A complete and uniform set of Spitzer photometric and spectroscopic 
observations are obtained for all stars in our dust disk evolution sample,
as described below. 
To derive statistically meaningful results on the disk and dust properties, 
we observe $\sim$50 stars in each of 6 logarithmically spaced age bins from
3 Myr (connecting our Legacy program to that of Evans et al.)
to 3 Gyr (beyond which there is strong emphasis by GTOs on debris disk
science; Beichman et al. 2005).  
Our targets span a narrow mass range (0.7-2.2~M$_\odot$)
and are proximate enough to enable a complete
census for circumstellar dust comparable to our model solar system
as a function of age. 
We measure the stellar photosphere 
at SNR$>$30 for $\lambda$ 3.6--24 $\mu$m, SNR $>$ 5 
at 70 $\mu$m (or SNR $>$ 5 for 5 $\times$ the current
zodiacal dust emission) with broadband photometry from 
IRAC and MIPS (subject to calibration uncertainties). 
To identify gaps in the dust distribution
created by the presence of giant planets from 0.2--10 AU, we require
relative spectrophotometry with SNR$>$30 from 5.3--11 $\mu$m, 
and with SNR $\sim$ 6--12 at wavelengths between 20--30 $\mu$m
with the IRS. 

A sub-sample of 35 stars comprises our gas disk evolution study 
with the high resolution mode of the IRS.  
This sample spans a range of spectral type (F3--K5), age (3--100 Myr),
activity (L$_x/L_{bol} \sim 10^{-3}-10^{-5}$), and a wide range
of infrared excess emission, with some preference for 
optically--thin excess in the mid--infrared. 
Fourteen stars were chosen for first-look observations and 
enable us to explore the limits implied by null results and guide our choice
of follow--up observations for additional stars drawn from our
dust disk survey.  

Our goal is to collect data capable
of realizing the fundamental limits imposed by instrument stability and 
systematic calibration uncertainties. 
Integration times are chosen according to each star's distance, age
and spectral type to reach uniform SNR at the photospheric 
limits, as specified above -- 
thereby providing a {\it complete census} of dust disks for our targets.

\subsection{Sample Selection}

The source list for FEPS consists of young near-solar analogs, stars 
ranging in mass from 0.7-2.2 M$_\odot$ though strongly 
peaked at 1 M$\odot$ (see Figure 2), 
and spanning ages 3 Myr to 3 Gyr (our Sun is 4.56 Gyr old). 
The stars are drawn from three recently assembled samples.

\begin{figure}
\insertplot{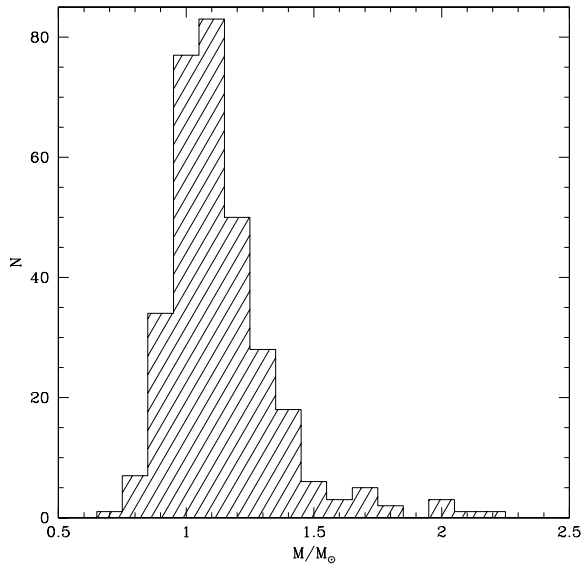}{3.0}{3.0}{2.70}{3.0}{0.4}{0}
\caption{
Distribution of masses for stars in the FEPS
sample.  The range spans 0.7--2.2 M$_{\odot}$ though it
is strongly peaked at 1.0 M$_{\odot}$.
}
\end{figure}

First, Soderblom et al. (1999) have produced a well-characterized set of 
$\sim$5000 solar-type stars spread over the entire sky 
(see also Henry et al. 1996) having parallaxes that place the stars 
within 60 pc, (B-V) colors between 0.52 and 0.81 (F8-K0 spectral types), 
and location in the Hertzsprung-Russell diagram within 1.0 mag of the 
solar-metallicity Zero-Age Main Sequence.  
This sample is fully complete out to 50 pc. The age distribution in such a
volume-limited region around the Sun is roughly flat in linear age units out to 
2.5 Gyr, at which point heating by the Galactic disk has increased the scale
height of older stars and thus removed them from the volume-limited sample.
From this catalog we have selected a sample with ages based on 
the R'$_{HK}$ chromospheric activity index from $\sim$100 Myr to 3 Gyr.  
However, being located more than 100 pc from the nearest sites of recent star
formation, the immediate solar neighborhood is deficient in stars with ages
younger than 100 Myr. Hence the volume limit was extended in 
order to identify large enough samples of young stars for the FEPS project.

We have conducted a new (e.g. Mamajek et al., 2002) and 
literature-based examination to identify stars  
whose ages are in the range 3-100 Myr.   
These were selected as having (B-V) colors between 0.58 and 1.15 or spectral 
types G0-K0, strong x-ray emission, kinematics appropriate for the young 
galactic disk, and high lithium abundance compared to the 120 Myr old Pleiades.
Young stars are copious coronal 
x-ray emitters and a large body of literature demonstrates the connection 
between x-ray emission, chromospheric activity, stellar rotation, and age. 
The surface density distribution of x-ray sources detected by the 
ROSAT all-sky survey reveals a concentration of objects coincident with 
Gould's Belt, a feature in the distant solar neighborhood (50-1000 pc)
comprised of an expanding ring of atomic and molecular gas of which nearly 
all star-forming regions within 1 kpc are a part.  These x-ray sources are 
thought to be the dispersed low-mass counterparts to a series of 1-100 Myr-old
open clusters and extant and fossil 
OB associations that delineate Gould's Belt (e.g. 
Torra et al. 2000; Guillout 1998).  Proper motion data enable us to select the
nearest of these young, x-ray-emitting stars with space motions consistent 
with those of higher mass stars having measured parallax, and hence estimate
their distances. 
Follow-up optical spectroscopy of these x-ray + proper motion selected stars 
is used to confirm youth and determine photospheric properties.  A total 
of $\sim$600 field stars are x-ray-selected candidate young stars. 

Finally, stars in nearby well-studied open clusters 
[IC 2602 (55 Myr), Alpha Per (90 Myr), Pleiades (125 Myr), Hyades (650 Myr)]
serve to ``benchmark'' our field star results
by providing samples nearly identical in age, composition, and birth
environment.  We considered all known members of these clusters meeting
our targetted mass / B-V color / spectral type range that were not part of 
GTO samples. 

From this large parent sample, stars were selected for potential observation 
with Spitzer if they met all of the following additional criteria.  The 
criteria were chosen to ensure sufficient signal-to-noise 
on the stellar photosphere out to 24 $\mu$m
with Spitzer and thus accurate characterization of the underlying 
photosphere both observationally and with stellar models.
\begin{itemize}
\item K $<$10 mag (young $<$100 Myr x-ray selected and cluster samples) 
or  K $<$6.75 mag (older 0.1-3 Gyr Hipparcos $+$ R'HK selected sample)
\item 24 $\mu$m background $<$1.70 mJy/arcsec$^2$ (x-ray selected samples) or 
$<$1.54 mJy/arcsec$^2$ (Hipparcos $+$ R'HK sample)
\item 70 $\mu$m background $<$0.76 mJy/arcsec$^2$ 
\item galactic latitude $|b| > 5^\circ$
(stars in IC 2602 were permitted to violate this criterion)
\item quality 2MASS JHK photometry, with no flags 
\item no projected 2MASS companions closer than 5" 
\item no projected 2MASS companions closer than 15" 
unless they are {\it both} bluer in J-K {\it and} fainter in K by $>$3 mag 
than the Spitzer target.
\end{itemize}

These critera were applied uniformly to our parent sample though
in the cases of a few exceptional stars (such as the IC 2602 sample)
some criteria were violated.
Next, targets appearing on Spitzer GTO 
programs were removed from the source list.  Also, to a limited degree, 
stars identified through spectroscopy or high resolution imaging 
literature published through March 2001 as being binary, 
with companions closer than 2", were removed.  These cases were all either 
spectroscopic binaries or visual binaries with small delta magnitudes and
the literature search was not exhaustive.  
Subsequent investigation using AO imaging have uncovered additional binary 
systems with larger delta magnitudes remaining within the FEPS sample
(e.g. Metchev \& Hillenbrand, 2004). 

Finally,
amongst the stars in our parent sample older than $\sim$600 Myr, 
approximately 1/2 were arbitrarily removed from our program in order to 
even out the age bins and bring the observing program
within the allocated number of Spitzer hours (350).

Based on pre-Spitzer
spectral energy distributions assembled from the literature, 2MASS, IRAS, ISO,
and ancillary observations conducted to date, several (5-7) of the youngest 
stars in our program show some hint of circumstellar material.  Because the
stars were randomly selected based on their kinematic and activity
properties as derived from optical information, these previously known dust
excesses do not bias the program.

To this source list a set of 14 stars was added, which were 
suspected to have optically thin dust excesses based on observations
from IRAS and ISO.  These stars formed part of our 
first-look gas disk detection survey and should not be used
to derive the statistics of dust debris as a function of age. 

\begin{figure}
\insertplot{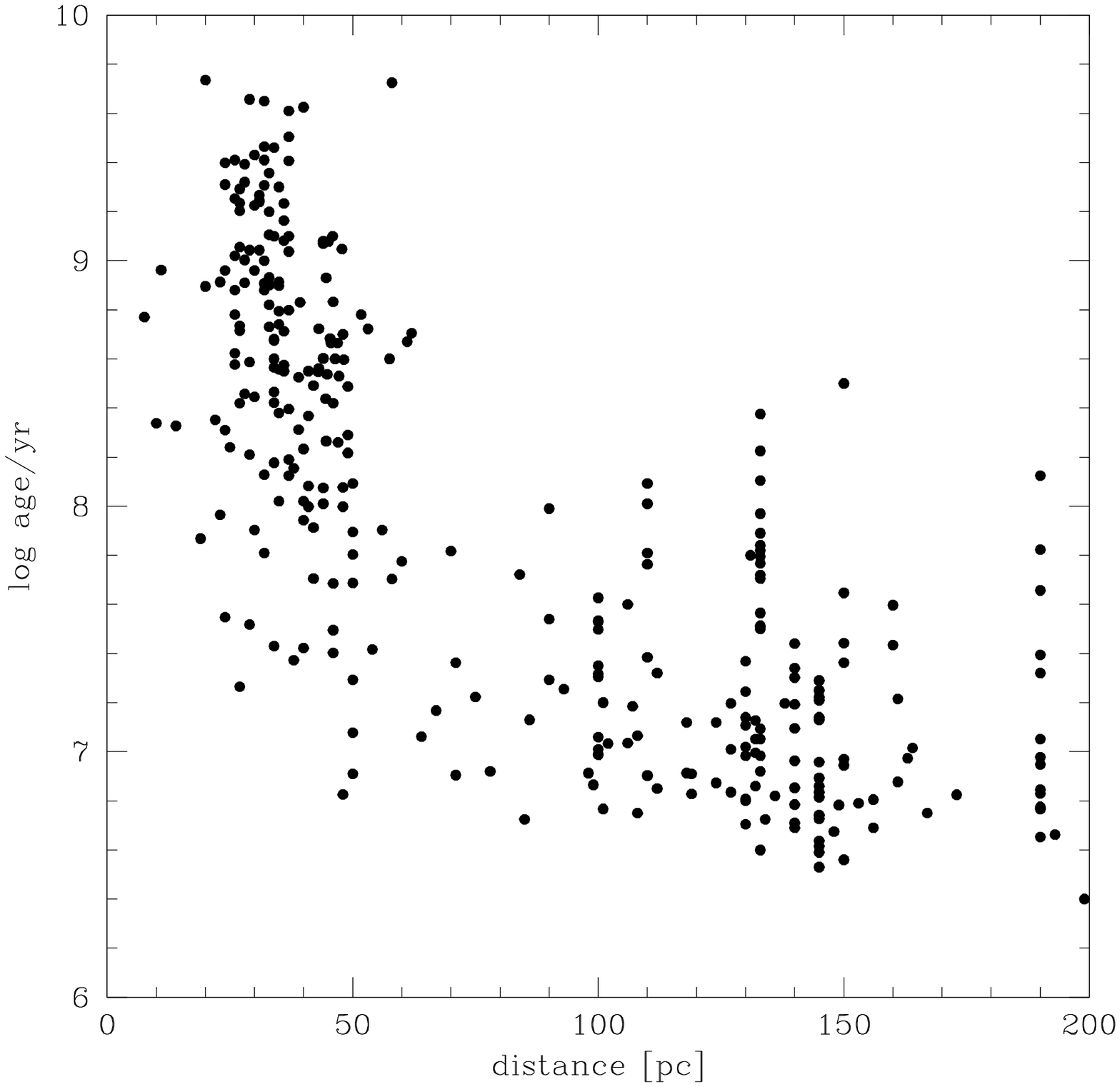}{3.5}{3.5}{2.70}{3.0}{0.4}{0}
\caption{
Estimated age versus distance for all
stars in the FEPS sample, comprised of nearby field
stars, open cluster members, and young association members.
Typical errors in age are less than 0.5 dex, while
typical errors in distance are less than 10 \% for stars within
50 parsecs, and within 30 \% for more distant targets.
}
\end{figure}

Our final target list for observations with Spitzer 
and ground-based ancillary programs consists of 328 solar-type stars distributed
uniformly in log-age between 3 Myr and 3 Gyr.  
Approximately 60 of these 
are members of the open clusters IC 2602, Alpha Per, Pleiades or Hyades.
The remainder are field stars distributed in distance between 11 and 180 pc.
The relation between distance and age for this sample is 
shown in Figure 3.  The complete source list
is presented in Table 1 (field stars), Table 2 (open cluster stars), 
Table 3 (young stars), and Table 4 (pre--selected IRS high resolution targets). 
The names, coordinates, and spectral types are those found in the
Spitzer Legacy Science Archive 
(http://ssc.spitzer.caltech.edu/legacy/all.html)
and details will be presented in Hillenbrand et al. (in prep).

\subsection{Spitzer Data}

\subsubsection{Astronomical Observing Requests}

FEPS uses all three science instruments on board Spitzer to
provide data from 3.6 $\mu$m to 70 $\mu$m (with a subset of the FEPS
stars also observed at 160 $\mu$m) including $\lambda \sim 7 - 38\mu$m
low resolution IRS spectra.  A detailed description of the observing
commands (Astronomical Observing Requests or AORs) that specify the
FEPS program can be found using the SSC's SPOT software to "View
Program" and entering Program Identification Number (PID) 148.  In
this section we outline the AOR strategy for each instrument.

Data are obtained with IRAC in three bands (3.6, 4.5, and 8.0 $\mu$m).
The first five FEPS objects observed as part of the early verification
program were observed in all four IRAC bands (including the
5.6 $\mu$m channel).  
FEPS stars are observed in the IRAC Subarray mode (32x32 pixels) at
frame-times of 0.02, 0.1, or 0.4 seconds, using the "4 point-random"
dither pattern at the medium dither scale.  At each of the 4 dither
positions, 64 images are taken at the same frame-time in each
band. Thus there are 256 images of each star for each IRAC band.

Low resolution (R $\approx$ 64-128) spectra covering the longer
wavelength ranges (7.4 -38 $\mu$m) in the SL1, LL1, LL2 modules of the
IRS are obtained for all FEPS objects.  Observations using the SL2
module (5.2 - 7.7 $\mu$m) were also obtained for the five validation
stars and for objects younger than 30 Myr.  High resolution (R $\approx$ 600) 
spectra are obtained for the 35 stars chosen for the gas detection experiment.
All IRS observations use standard starring mode, and high accuracy peak-up.

MIPS photometric imaging data at 24 and 70 $\mu$m are obtained for
the entire FEPS sample.  We achieve SNR $>$30 at $\lambda\leq24\mu$m
for our entire sample and SNR $>$5 at the photospheric limit for
as much of our sample as practical at $\lambda=70\mu$m.  For sources
whose photospheres we are unable to detect in a reasonable time 
(10 cycles), we
achieve SNR $>$ 5 on dust debris at $\times$ 5 current Solar System values. 
  MIPS 160 $\mu$m data are also obtained for 10\%
of the lowest background targets, chosen to span the age range of the
full sample.  The MIPS default scale photometric mode acquires data in
multiple pointings with small offsets between each pointing to
alleviate instrumental artifacts and cosmic rays.  Due to the
stability of the Si:As array, these multiple pointings for the 24
$\mu$m data allow repeatability to be used as an accurate estimate of
internal measurement (precision) uncertainty.  For 70 and 160 $\mu$m,
the multiple pointings are used to calibrate the time-dependent
detector response, and therefore only a final mosaic image is produced by the
photometry mode for the Ge:Ga arrays.

\subsubsection{Data Reduction}
The SSC pipelined data products from FEPS can be accessed through the
SSC's LEOPARD archive browser.  In this section we outline the general
data reduction strategy for each instrument.  We refer the reader to
FEPS data publications for more detailed description of the data
reduction applied to individual sources (see also the FEPS
Explanatory Supplements that accompany our data releases through 
the Spitzer Science Center). 

\subsubsubsection{IRAC}
IRAC data frames are processed through the SSC pipeline to produce
Basic Calibrated Data (BCD) images.  Sixty four images are obtained at
each of the four dither positions for a total of 256 images of the
object, in each band.

Flux densities are derived using aperture photometry.  A Gaussian fit
to the PSF is used to center the target aperture.  The target and
background annuli are optimized for the ensemble of observations to
maximize the measured SNR. These measured flux densities are then
corrected to the standard calibration aperture sizes supplied by the
SSC. To estimate measurement uncertainties we assign the standard
deviation of the 256 measurements as the one-sigma (1$\sigma$)
internal uncertainty.  The total uncertainty reported by FEPS is the
(square-) root of the sum of squares (RSS) of this internal
measurement uncertainty and the calibration uncertainty as published
by the SSC. Although the calibration of IRAC assumes a flat
($\nu$F$_{\nu} = const.$) spectrum across the band-pass, no color
corrections have been applied to the FEPS IRAC measurements.  The
prescription for color-corrections is presented in the IRAC Data
Handbook available through the SSC.

\subsubsubsection{IRS}

IRS data are first processed through the SSC pipeline to remove
instrumental artifacts including dark current, droop signal and flat
field structures.  From these pipelined data we proceed with the
intermediate "droopres" data product.  The SMART reduction package
developed by the IRS Instrument Team at Cornell (Higdon et al 2004) is
then used to extract the spectra.  As a first step, we correct for the
background emission and stray-light by subtracting the images obtained
from the two slit positions at which a object is observed for each
module and order (automatically in standard starring mode).  This
results in a set of images containing a positive and negative spectrum
in each observed order.  Before extraction, all hot or dead pixels in
each image are replaced.

For the spectral extraction we use a straight-sided (boxcar) aperture
limiting the extraction area around the positive source in the
background corrected images.  Since all observations in the FEPS
legacy program use high accuracy peak-up and the pointing of 
Spitzer is good to within 0.4" radius (1$\sigma$), we fix the
position of the aperture in each spectral order.  The width of each
aperture is chosen such that 99\% of the source flux is within the
aperture for all wavelengths in the order.  A mean spectrum over all
slit positions and cycles is computed for each individual order from
the spectrum for each nod position and cycle.  The orders are then
combined.  The quoted uncertainties are the 1$\sigma$ standard
deviation of the distribution of data points used to calculate the
mean spectrum over all cycles and nod positions, modified to include
the errors of the photometric calibration.  Finally, the orders are
stitched together with unit weight by taking the mean flux at
overlapping wavelength points.

\subsubsubsection{MIPS}

MIPS data were originally reduced using the Data Analysis Tool
developed by the MIPS Instrument Team at the University of Arizona
(Gordon et al.  2004), since the MIPS Instrument team was charged with
fast development of the primary redcution algorithms.  This package
uses the raw data product available from the SSC data archive.  Dark
subtraction, scan mirror dependent flat field, electronic nonlinearity
correction, droop subtraction, and cosmic ray rejection are applied.
For the final FEPS releases, the SSC pipeline products are used, since
the combined efforts of the Instrument Team at the University of
Arizona have been successfully implemented at the SSC and the two
reduction schemes have converged on a common, validated product.

Flux densities for each band are derived from aperture photometry.
The position of the aperture was found by fitting a two-dimensional
Gaussian to the core of the PSF when the object is detected.  For
non-detections, as is often the case for 70 and 160 $\mu$m, the
aperture is centered on the object coordinates.  
Aperture correction factors are applied to match the ``infinite
aperture calibration'' defined by the MIPS Instrument Team (MIPS Data
Handbook).

We use the standard deviation of the photometry from the stack of
individual dither images (24$\mu$m), or the RMS noise in the
background propagated over the pixels in the object aperture as
estimates of the random photometric uncertainty (70 and 160$\mu$m).
Total uncertainties are the RSS of the internal uncertainties and the
published calibration uncertainties.  As for IRAC, color-corrections
are not applied.  However, we note that the MIPS team assumes a 
10,000K black-body for its calibration (MIPS Data Handbook).

\subsubsection{Verification and Validation}

Quality control is applied to the Spitzer observing program as
follows.  All observations are verified, meaning checks on whether the
correct source was observed, in the requested instrument mode, and
following the prescribed AOR. The observations are further validated
by considering photometric uncertainties derived from the observations
compared with theoretical uncertainties based upon expected photon count rates
and other known noise sources such as extragalactic confusion at 70
and 160$\mu$m.  Expected versus derived SNRs are assessed for
observations at wavelengths where the measured flux densities are
consistent with being photospheric.  Comparison with photospheric
model expectations enable us to investigate systematic offsets in 
expected versus observed fluxes as a function of source color and
brightness, although we are unable to unambiguously 
separate errors in the models from errors in the data. 
For wavelengths $\lesssim
24\mu$m, the exposure times were sufficient to measure the
photospheric emission with expected SNR $> 30$.  
We also verify that fluxes derived from different instruments
over common wavelength ranges agree within the errors. 
Further details can be found in the Explanatory Supplements
that accompany our data release through the Spitzer Science Center. 

\subsection{Ancillary Data}

In addition to the Spitzer observations, we are engaged in a 
rich ancillary observing program that both complements and aids
our interpretation of the Spitzer spectrophotometry.  These data include
ground-based 10$\mu$m, sub-millimeter, and millimeter photometry,
an echelle spectroscopic survey for photospheric characterization, and
an adaptive optics imaging survey for companion detection and characterization.

{\it Dust Mass Constraints:}
We have engaged in a limited ground-based mid-infrared campaign on a few tens
of FEPS targets.  Imaging photometry 
at 10 $\mu$m of selected members of the FEPS sample 
were obtained with LWS on the Keck I telescope and SpectroCam-10 (SC-10) on the 
Hale 5-meter telescope (Metchev et al. 2004) and also with
MIRAC3 on the Magellan I telescope (Mamajek et al. 2004).  The observations
were largely consistent with photospheric emission with few exceptions where
excess emission was detected, indicative of terrestrial zone dust. 
We also searched for dust located at larger radii and hence
too cold to radiate strongly in the MIPS 160 $\mu$m band. We
obtained sub--mm and/or mm continuum observations for approximately 1/3
of our sample. 
Millimeter observations were obtained using the Owens Valley Radio Observatory 
(OVRO) millimeter-wave interferometer at 3.1 mm or the  37-element SIMBA 
bolometer camera on the 15m Swedish-ESO Submillimetre Telescope (SEST)  
at 1.2 mm, for a total of 89 stars.  Submillimeter 
observations at 350 $\mu$m were obtained for 6 stars using the SHARC 
bolometer camera on the 10.4m telescope of the Caltech Submillimeter 
Observatory (CSO).
These observations are discussed in detail by Carpenter et al. (2005).  
Thirteen FEPS sources were observed by Najita and Williams (2005) at the 
James--Clerk--Maxwell Telescope (JCMT) 
using SCUBA.  The sources were selected with an emphasis on those that
are young (10 Myr to 300 Myr) and nearby ($<$ 50 pc).  Three sources
were detected including HD107146 (Williams et al. 2004).  

{\it Gas Mass Constraints:}
As a complement to our Spitzer H$_2$ program we are attempting
CO rotational transition detection at millimeter wavelengths with OVRO,
Sub--millimeter Telescope facility of the Arizona Radio Observatories, 
and the JCMT, as well as 
fundamental vibrational emission at mid-infrared 
wavelengths (4.5 $\mu$m), and the pure rotational 
transitions of H$_2$ at 17 $\mu$m.  A sample of $\sim$ 20 sources
were observed in the CO(2--1) line with the SMT and
upper limits were derived.  These data are being combined
with Spitzer observations and results are reported in 
Pascucci et al. (2006). 
In addition, Najita and Williams (2005) have searched for
CO(3-2) emission from two of the submillimeter excess sources and place
limits on the gas mass in these systems.  In the case of HD107146, a
conventional analysis suggests that the upper limit on the gas-to-dust
ratio is much less than primordial.  

{\it Age Diagnostics:}
An important aspect of our program is determination of the tightest constraints
possible on the ages of our sources (Hillenbrand et al. in preparation). 
To do so we consider a number of
diagnostics related to activity, which generally decreases with increasing
stellar age, or elemental abundances, in 
particular Li I.  We have assembled all x-ray information from the ROSAT
archives.  We have R'$_{HK}$ indices for over 3/4 of our sample, from the 
literature or newly measured from our ancillary high dispersion 
(R$\approx$20,000 from 3600-9500$\AA$) optical spectra. In addition to 
Ca II H\&K core emission line strengths, we are also measuring
H$\alpha$ emission/absorption equivalent widths, Li I equivalent widths,
and rotational velocities, all of which change with stellar age.  
A full discussion of stellar age indicators and their likely
uncertainties is beyond the scope of this paper (cf. Hillenbrand et al.).  
However, we find generally that when multiple age indicators are available for a
given star they agree with one another to better than 0.5 dex in log--age.  
This level of accuracy is adequate for investigating general trends
in debris disk evolution.  Finally, we can also 
derive effective temperatures, gravities, and metallicities 
for each star through spectral synthesis modelling. 

{\it Stellar, Sub-stellar, and Planetary-mass Companions:}
In order to place our own Solar System fully in context, we must 
consider the effects of stellar multiplicity at the same time we are
considering dust disk evolution. Our Sun's 
planets exist at orbital radii ranging from 0.4-30 AU with the Kuiper Belt 
extending out to at least 50 AU, and our Sun is not  a member of a multiple
star system.  A significant fraction (30-80\%) of all sun--like 
stars do appear to be 
born in multiple star systems (binaries, triples, quadruples; Mathieu et al. 
2000) and, as shown for solar-type, solar-neighborhood multiples by Duquennoy 
\& Mayor (1991), the distribution of orbital periods is lognormal and
peaked at 180 days or 30 AU -- i.e. within our current Solar System.
Searches for companions (stellar, sub-stellar, and planetary-mass)
to members of our Spitzer sample
via both high resolution imaging and spectroscopy are underway, in order
to assess the role of multiplicity in disk evolution.

With the adaptive optics (AO) at the Palomar 200" 
we have observed every northern star on our program with short JHK$_s$ 
exposures, designed to detect bright companions as close as 0.1" 
(5 AU for sources at 50 pc).  Such companions within the Spitzer beam
(5" at 24 $\mu$m) are critical to account for when analyzing 
spectral energy distributions.  For a selected subset of our stars we are also 
performing deep K$_s$-band AO coronography designed to detect much fainter 
companions.  Due to evolution of the mass-luminosity relationship
and contrast limit, 
there is an intricate grid of tradeoffs in the companion mass 
detection limit as a function of system age and orbital separation, with
sensitivity to lower masses achievable at younger ages and larger
separations.  In the case of our target list, we are sensitive to masses
as low as 3-10 M$_{Jupiter}$  (for example, at separations of 2"
to $\Delta K$= 13 mag at SNR = 5). Follow-up proper motion, photometry, and
spectroscopy is conducted with the Palomar 200" and with Keck
(e.g. Metchev \& Hillenbrand, 2004; 2006, in press). 

High dispersion spectroscopy is also being used to identify companions.
Several double-line or single-line spectroscopic binaries have been 
found from our Palomar 60" echelle spectroscopy (White et al., 
submitted).  
Further, at least 25\%
of our Spitzer target stars are located on various radial velocity
planet search programs, a number of which are already known to have
planetary companions or will be found with planetary companions over the
next decade.  Several FEPS targets are also being monitored photometrically
in order to derive rotation periods from star--spot activity on 
the stellar surface. 

\section{Supporting Theoretical Framework}

Physical theories are needed to guide our observational program and to help 
interpret the results.  The basic problem is to understand the formation and
evolution of planetary systems based on observed spectral energy distributions,
and spectroscopic observations.   In support of this work, we 
have undertaken a limited modeling effort aimed at constraining: 
1) basic disk properties utilizing a minimum of assumptions; 2) 
the amount of remnant gas in disks based on IRS high resolution
spectra; and 3) the diversity of planetary architectures based on 
estimates of geometric dust distributions derived from the SEDs. 

\subsection{Toy Model for Solar System Evolution}

\begin{figure*}
\epsscale{2.0}
\plottwo{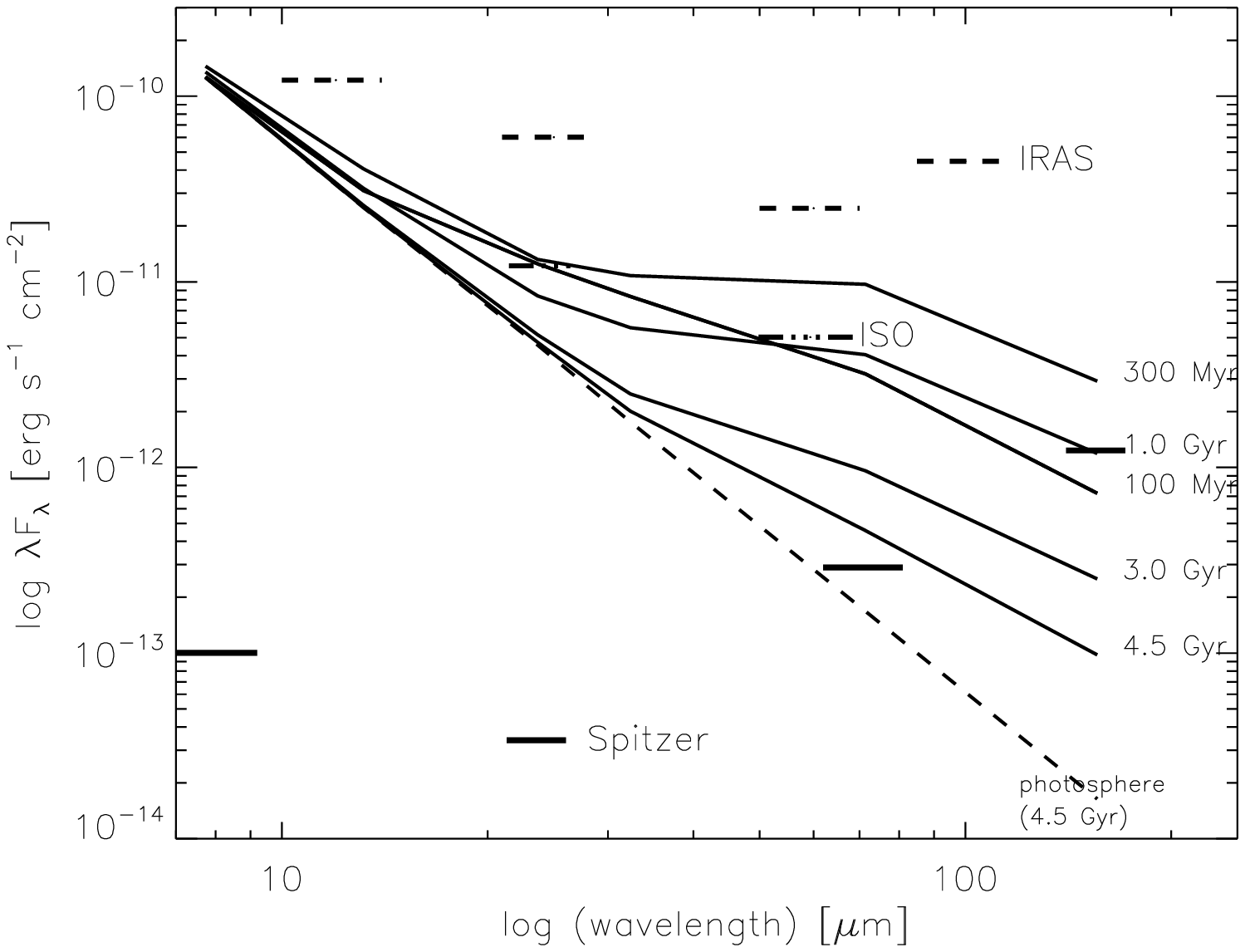}{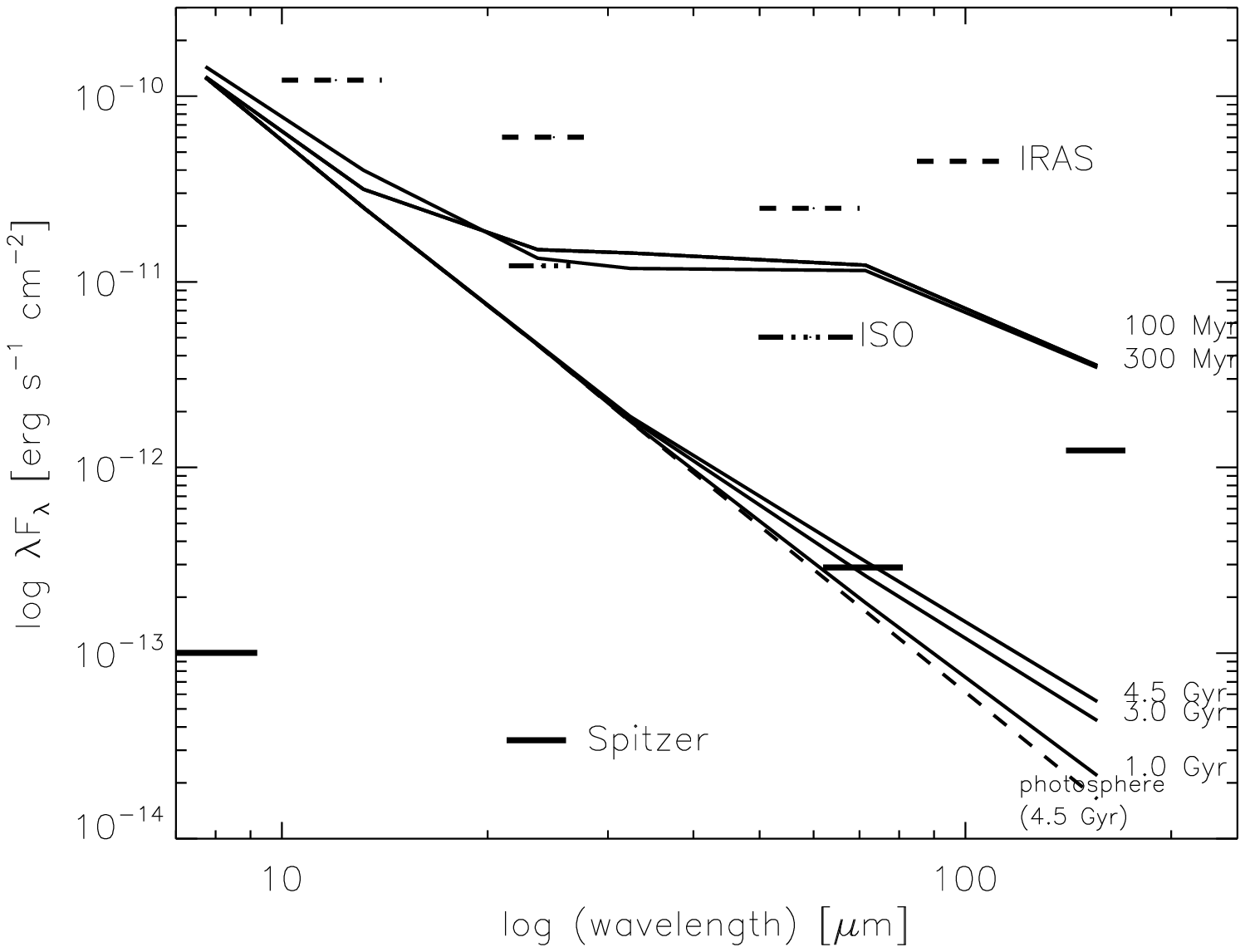}
\caption{\small 
Toy model for the evolution of our Solar System
debris disk surrounding the Sun as observed from a distance of 30 parsecs.
The model shown in the top panel
includes continuous removal of planetesimals starting
from the minimum mass solar nebula in solids, and evolving towards
the present day, without any dramatic clearing event such as
the Late Heavy Bombardment.  The model shown in the bottom panel
begins with the minimum mass solar nebula in solids
as in the top panel, but includes a dramatic clearing event such as
the Late Heavy Bombardment between 300 Myr and 1 Gyr in accordance
with recent models of Gomes et al. (2005; see also Strom et al. 2005).
}
\end{figure*}

We start with a basic model of the evolution of our own solar
system.  As is well known, our planetary system contains two 
major debris belts: the Kuiper Belt and the Asteroid Belt, both of 
which are generating dust through mutual collisions of larger 
parent bodies.  Figure 4 shows the dust mass and the observed 
SEDs predicted by two plausible models 
 for the evolution of our Solar System from 100 Myr to 4.5 Gyr,
 as viewed from 30 pc. The models 
 assumes only a minimum mass solar nebula and
 planetesimal scattering and collision frequencies according to
 two simple analytic representations:  one (bottom) including the effects
 of dynamical instability postulated to have removed a large fraction
 of dust--producing parent bodies in the asteroid and Kuiper Belts
 in our Solar System (Gomes et al. 2005; see also Strom et al. 2005) 
and the other (top) excluding those
 effects.   In the former case, dust production diminishes linearly
  in time as expected for a high density planetestimal belt in
  collisional equilibrium where dust is ultimately removed through
radiation pressure blowout of the smallest fragments
(Dominik and Decin, 2003; Wyatt, 2005) until the
instability occurs at approximately 500--600 Myr.  In the latter
case, the belt grinds itself down to the low density limit
where dust removal is dominated by P--R drag and the dust mass
observed decays at t$^{-2}$.
  In the absence of this instability, our hypothetical solar system
 is detectable by Spitzer with IRAC (zodiacal dust disk) and MIPS
 (Kuiper disk) from 30 pc at age 100 Myr, while only the Kuiper belt
 dust would be found at an age of 4.5 Gyr.
In this way, we can attempt
to address whether or not our Solar System is common or rare
compared to typical stars in the disk of the Milky Way galaxy.
Details concerning this model can be found in Meyer et al. 
(2006).

One deficiency in this model is neglecting the drag on orbiting grains 
due to stellar winds.  Azimuthal wind drag could dominate radiation P-R drag 
for the high mass-loss rates expected from young solar-type stars (Jura 2004).
This would decrease grain lifetimes in systems that are P--R 
drag dominated, and diminish the number of systems thought to be collisionally
dominated.  However, recent work (Wood et al. 2005) suggests that the large effects
suggested in earlier work (Wood et al. 2002) may have been overestimated.
While this effect can be important, its magnitude is still uncertain. 

\subsection{Constraints on Dust Properties} 

The observed SEDs from dust disks are determined in part, by the 
optical properties of the dust contained therein.  Both particle
size and composition are important in determining the absorption
and emission properties of the dust, thus determining its temperature
for a given distance from the central star.  In the absence of 
constraints on dust properties from spectral features, resolved
images of disks in thermal emission or scattered light, 
or far--IR/sub--mm spectral slopes, models to explain the 
observed SEDs of debris disks ar necessarily degenerate
(see for example Hines et al. 2006).  Spectroscopic observations
of solid state features can provide important constraints on physical 
models for the dust.  For example, large particles 
($a_{DUST} > \lambda_{resonance feature} / \pi$) are
not efficient radiators at their natural resonance 
frequencies.  Thus the absence of expected solid state features
from abundant species can indicate a minimize grain size. 
Similarly, specific chemical compositions of dust can change
the shape and central wavelength of resonance features, or
indicate significant structural differences in the dust
(e.g. amorphous versus crystalline silicates).  Discerning
the difference between Mg-- and Fe--rich end member silicates
and fractions of amorphous to crystalline silicates provide
crucial information concerning the chemical properties of
the nebula in which parent body planetesimals formed, 
as well as the processing history of the dust. 
Detailed models exploring these effects 
are described in Wolf \& Hillenbrand (2003) and Bouwman et al. 
(in preparation). 

\subsection{Dynamics of Disk--Planet Interactions} 

As part of our theory effort we have developed numerical
tools to model debris disks originating from an outer belt of planetesimals
and evolving under the effect of gravitational perturbation from giant
planets in various planetary configurations and for different dust
particle sizes and compositions (Moro-Mart\'{\i}n \& Malhotra, 2002; 
Moro-Mart\'{\i}n \& Malhotra, 2003; and Moro-Mart\'{\i}n  et al. 2005). 
Even though the particle dynamics is chaotic, our method can robustly estimate the
equilibrium radial density distribution of dust. The dust density structure carved
by giant planets affects the shape of the disk SED, in a manner that depends upon the
the mass and location of the perturbing planet as well as the grain properties
(chemical composition, density and size distribution). We found that the SED of a
debris disk with embedded giant planets is fundamentally different from that of a
disk without planets, the former showing a significant decrease of the near/mid-IR
flux due to the clearing of dust inside the planet's orbit. The SED is particularly
sensitive to the location of the planet, i.e. to the area interior to the planet's
orbit that is depleted in dust due to gravitational scattering by the planet. Our
dynamical models show that for a planet in a circular orbit with semimajor
$\it{a}$$_{pl}$, the radius of the depleted inner zone is in the range of
0.8--1.2$\times a_{pl}$ depending upon the planet mass. Our models also show that
the dust depletion factor (i.e.~the ratio between the dust density inside and
outside the depleted region) depends significantly on the planet mass when the
mass is in the range $1M_{Nep} < M_{pl} < 3M_{Jup}$. However, there are two
issues that complicate the interpretation of the SED in terms of planet location:
(1) The SEDs are degenerate. In particular, there is a degeneracy between the dust
grain chemical composition and the semimajor axis of the planet responsible for
inner the gap. For example, the SED of a dust disk dominated by weakly absorbing
grains (Fe-poor silicate) has its minimum at wavelengths longer
than those of a disk dominated by strongly absorbing grains (e.g. carbonaceous and
Fe-rich silicate), which may be mistaken by the presence of a larger inner gap.
This degeneracy can be resolved either with high resolution 
spectroscopy (which would
constrain the grain chemical composition as discussed above), 
or high spatial resolution images (which
would spatially resolve the inner edge of the dust disk).
(2) Because of the Spitzer sensitivity limit, the debris disks observed
by FEPS may be in the collision-dominated regime, where the dynamics of the dust
particles are dominated by collisions rather than P-R drag 
(Dominik \& Decin, 2003; Wyatt, 2005).
This may result in the dust particles suffering multiple collisions, that could break them down into
smaller and smaller grains until they are blown out from the system by radiation
pressure, before they have time to migrate from the dust-producing plantesimals to
a planet-crossing orbit.

\subsection{Models of Gas in Disks} 

In order to interpret our high resolution spectroscopic observations
of gas in disks from ages of 3 Myr to 100 Myr, 
we have developed detailed thermo-chemical models of gas and dust in
optically thin disks (Gorti \& Hollenbach 2004). Our
models calculate the gas spectral line emission and dust continuum emission
for comparison with observed data and thereby infer disk properties.
The models calculate the gas and dust temperatures separately, assuming
a balance between the heating and cooling processes. We
include various heating sources for the gas, such as collisions with warm
dust heated by stellar radiation, X-rays, exothermic chemical
and photo-reactions, cosmic rays and grain photoelectric heating.
The gas cools by ionic, atomic and molecular line emission. The
disk  temperature structure, vertical density structure, and  chemistry
is self-consistently
calculated in our disk models by requiring thermal balance,
vertical pressure equilibrium and by assuming steady-state chemistry.
 Our chemical network consists of 73 ionic,
atomic and molecular species involving H, He, C, O, Si, Mg, Fe and S and 537
reactions. Inputs to our models are stellar parameters such as the radiation
field at X-ray, UV and visible wavelengths, the disk surface density
distribution in gas and dust,
and dust properties (chemical composition and size distribution). Most of
the stellar and dust parameters are determined through ancillary
observations and by modeling of the
dust continuum observed through the FEPS program. For a given dust and gas
surface density distribution, our theoretical models can predict the
spectral line emission from
the gas and the dust continuum emission for comparison to observations.

We are in the process of developing similar gas disk models for younger,
optically thick dust disks. We use a two-layer model for the dust
temperature calculation (Chiang \& Goldreich 1997), and adopt a procedure
similar to that for optically thin disks for the gas temperature. The
gas and dust temperature are calculated separately and the emergent
line plus continuum spectrum computed.

 Spitzer IRS in the high resolution mode is capable of detecting
(3$\sigma$) 
warm gas ($T_{gas} \sim 100$ K) masses of $\lesssim 0.2 M_J$ in disks
at distances of 160 pc or less (Gorti \& Hollenbach 2004).
The FEPS H$_2$ program will measure spectral line fluxes (or upper
limits to line fluxes) and these will be compared with the gas models
to infer the gas masses in disks and the spatial distribution of gas
(e.g. Hollenbach et al. 2005; Pascucci et al. 2006). 

\section{Analysis Plan}

Our approach to analyzing the data collected as part of FEPS
starts with as few assumptions concerning the physical nature
of the observed system as possible, and proceeds to more
ellaborate models concerning the excess emission.  At each
stage, additional assumptions are made which enable a richer
interpretation of the data.  However, the certainty of our
conclusions diminishes as we proceed.  In any event, we attempt
to clearly state our assumptions as we go, and try to be careful
not to proceed on the basis of assumptions that are demonstrably
false.  In following this analysis proceedure, it is important
to always keep in mind how the results depend on the input
assumptions, as well as explore the full range of model parameters
allowed by the data, including the uncertainties.

Emission in excess of that expected from the stellar photospheres is
found by subtracting model photospheric flux estimates from the
observed fluxes across the wavelength range accessible to Spitzer.
We execute the following generic analysis for all sources with
excess emission of 5 $\sigma$ or greater at one or more wavelengths,
or equivalent detections of lower SNR but at two or more wavlengths.
We begin by calculating an approximate dust temperature if the
excess is detected at two or more wavelengths.  If the excess is
detected at only one wavelength, we derive a temperature limit using the excess
and the bluest broadband point without excess. 
With this temperature fit to the data, we then estimate
the ratio of excess luminosity in the infrared to the total
stellar luminosity $f \sim f_{IRX} / f_*$.

If the source
presents an excess over a broad range of wavelength, we explore
whether a range of dust temperatures would be a more appropriate
model.  Emission that appears to be optically--thick in the
direction perpendicular to the disk from the point of view
of the observer over a range of wavelength is initially assumed
to be a primordial gas--rich disk left over from the formation
of the star.  If high resolution
IRS data are available for the source we can assess whether
those these observations constrain the amount of molecular gas
remnant in the disk.  Other observations such as mm--wave CO
data or evidence for active accretion onto the star can also
suggest a primordial disk.  Evidence for a flared disk
geometry from the SED can also provide evidence for a gas--rich
primordial disk.

If there is no evidence for remnant gas in the disk, we proceed
under the premise that the disk might be a debris disk, where
the dust we see is generated through collisions of larger parent
bodies in a planesimal belt.  Assuming the grains are large,
efficient absorbers and emitters of light, we can calculate
the required dust cross--sectional area for the emitting grains,
and determine a plausible radius in the disk for the planetesimal
belt.  For an excess observed at a wavelength $\lambda$, we
can assume that grains larger than $\lambda / \pi$ can be
treated approximately as perfect blackbodies.
Given the luminosity of the star, we can also calculate
the radiation pressure blow--out size, providing a bound
on the smallest particles that could be responsible for the
radiation.  To further constrain the grain size distribution
and composition, we also search for evidence of solid state
emisson features in the IRS low resolution spectra obtained
for each source.  If grains exist in the circumstellar
disk at temperatures corresponding to emission at the
appropriate wavelength (e.g. 300 K for the 10 $\mu$m silicate
feature), the lack of expected spectral features attributed
to specific grain compositions can place constraints on their
abundance (e.g. upper limit to the fraction of crystalline
silicates in the debris dust) or particle size (grains
larger than $\lambda_{solid state feature} / \pi$ will not
be efficient radiators in the reasonance feature) or both.
These observations, as well as the observed spectral
slope in the far--infrared/sub-mm, constrain the grain size
distribution, and thus the range of plausible models for
the radii in the disk responsible for the emission at a
given temperature.  For example, very small grains ($<$ 0.1 $\mu$m)
can reach the same temperature at $>$ 30 AU as very large
grains ($>$ 10 $\mu$m) at $<$ 3 AU (Backman \& Paresce, 1993).

Armed with this information, we can calculate basic quantities
such as the mass surface density of emitting grains for a given
particle size/radius in the disk.  This enables us to compare
the lifetime of grains of various size under the assumption
that they are subject to both mutual collisions in the disk
as well as the effects of P--R drag (e.g. Burns et al. 1979).
In most cases, observed debris disks have surface densities
so high that many collisions will occur between dust grains
before they have time to evolve in radius significantly
under P--R drag (Dominik \& Decin, 2003; Wyatt, 2005).  In this case,
the grains are eroded down to the blowout
size and removed from the system through radiation pressure. 
If the IR excess is confined
to cooler temperatures (and therefore larger 
radii), we can also place limits on the mass surface density
{\it inside} of R$_{inner}$ and characterize the presence of
an inner hole in the dust distribution.  In principle,
limits on the mass surface density outside of the observed
disk radius and R$_{outer}$ could be constrained by FIR
and sub-mm observations as well.

In the case where the IR excess is detected with SNR $>$ 5
at several wavelengths, we explore disk model parameters
in a robust way by modelling the excess emission and calculating
the best--fit by computing the reduced chi--squared metric.
Given the number of degrees of freedom in the model, the number
of observations, and robust uncertainties in the observations,
the reduced chi--squared can provide an estimate of the
probability that any parameter lies within a range of 
values (confidence interval) if the data were drawn from a particular 
model.  As a result we
can define {\it contours} of allowed (correlated) parameters
in a multi--dimensional space defined by the model parameters.
This requires clearly defined model parameters that can be
varied over a range to produce acceptable fits to the data,
and clearly defined uncertainties in our data that can be
interpreted as errors in a gaussian probablistic sense.

Even in the absence of robust parameter estimation described
above, we can often constrain the family of permitted models
or offer a preferred model on the basis of likely physical
scenarios and Occam's razor.  For example, in the case where two models match
well the observed excess (small grains at large radii vs.
large grains at smaller radii), one can argue that the large
grain model might be preferred if the surface density of that
model were so large as to enable the disk to maintain an inner
hole on the basis of mutual collisions down to the blow--out
size.  The small grain at large radii model might have a
surface density so low that an interior planet might be
required in order to avoid grains filling in the hole
under the action of P--R drag.  Thus the large grain model
at small radii requires fewer assumptions and might be
preferred.  Furthermore, we know that T Tauri stars in their
youth have optically--thick circumstellar disks from $<$ 0.1
AU that extend to $>$ 30 AU.  It might seem implausible to
require a massive circumstellar disk composed entirely of
small grains $<$ 1 $\mu$m at R $>$ 100 AU with no evidence
for dust or planetesimals inside of 30 AU, rather than a
more modest remnant disk composed of larger grains at radii
where we know giant planets form and disks exist around
sun--like stars (cf. Kim et al., 2005; Hines et al. 2006).

Finally, all of the models considered should make specific
predictions that can be tested with follow--up observations.
For example, scattered light imaging with adaptive optics
on large ground--based telescopes, or utilizing coronagraphy
on the Hubble Space Telescope can provide constraints on the
radial extent of small grains in the debris disk systems.
More sensitive Spitzer observations at mid-- and far--IR
wavelenghts can improve upon low SNR initial detections.
Sub--mm observations can constrain disk models, particularly
the use of interferometers to resolve the FIR/sub--mm
emission initially detected by Spitzer.  In general, having
resolved images of disks and one or more wavelengths in
thermal emission, or in scattered light break many of the
degeneracies associated with SED modelling. 
For the nearest, youngest systems where a hole in the
dust distribution is inferred from SED modelling, we
can test those predictions using high contrast imaging
techniques to search for warm massive planets in the
circumstellar environment R $>$ 5 AU.   The connection
of dust disk emission with the presence/absence of
giant planets is still poorly understood
(Moro--Martin et al. in press; Beichman et al. 2005)
and one of the key goals of the FEPS science program.

\section{FEPS Data Products} 

As part of our commitment to the Spitzer Legacy Science Program, 
we plan to deliver high quality enhanced data products for the 
benefit of the community, as well as documentation that will 
enable archival researchers to utilize these data in the most
efficient way possible.  In addition to ground--based ancillary 
data, and the Spitzer database itself, we also provide resources
that enable careful scrutiny of the Spitzer calibration, and
tools for the research community interested in interpreting observations
of debris disk systems. 

\subsection{Ancillary Data Products}

Ancillary data collected in support of FEPS are provided via 
the published literature and include all data discussed above:

\begin{itemize}
\item
10 $\mu$m photometry (e.g. Metchev et al. 2004; Mamajek et al. 2004). 
\item
Sub-millimeter and millimeter continuum photometry (e.g. Williams et al. 
2004; Carpenter et al.  2005; Najita \& Williams, 2005). 
\item
CO gas line measurements or upper limits (e.g. Najita \& Williams, 2005; 
Pascucci et al. 2006). 
\item
Stellar age indices (e.g. Hillenbrand et al. in preparation). 
\item
Stellar photospheric properties (provided through Legacy 
deliveries to the Spitzer Science Center; see section 6.3). 
\item
Detected companion properties and limits on undetected companions
(e.g. Metchev and Hillenbrand, 2004; 2006). 
\end{itemize}

Selected data tables from these sources 
are also included in the Spitzer Legacy Science archive
(particularly the mid--infrared and sub--millimeter survey data 
and those data used in fitting photospheric parameters). 

\subsection{Spitzer Data Products}

Spitzer data are provided to the SSC for all 328 stars in the
FEPS sample as raw and SSC pipelined products (accessed through
LEOPERD) and as ``enhanced data products''  
(http://ssc.spitzer.caltech.edu/legacy/all.html). 
A Pointed Observations
Photometric Catalog (POPC) is available consisting of flux densities
for IRAC (3.6, 4.5 and 8.0$\mu$m) and MIPS (24 and 70 $\mu$m)
observations for all sources.  IRAC 5.8 $\mu$m flux densities are
available for five stars that were part of the early validation
portion of the program.  Due to time constraints imposed by slight
modifications in the expected SNR based on the updated on--orbit
performance of the instruments, we chose to drop the 5.8 $\mu$m band
from our general survey rather than decrease the number of targets. 
\footnote{Because FEPS utilized the sub--array mode of IRAC
observations reserved for bright stars, we do not obtain 5.8 $\mu$m
observations simultaneously with the 3.6 $\mu$m observations as is the
usual case.} Similarly, 160 $\mu$m observations are available for
approximately 10\% of the FEPS sample.  The confusion limit at 160
$\mu$m for most of our targets was well above the sensitivity level
needed to test our toy model for the evolution of our Solar System
through observation of sun--like stars from 3 Myr to 3 Gyr.
Therefore, we chose a limited campaign of 160 $\mu$m observations in
order to preserve the ``discovery space'' ($> \times$ 30 compared to
ISO at these wavelengths) enabled by MIPS 160 $\mu$m observations for
a random subset of our sample.  We also include an image atlas based
on mosaicked IRAC and MIPS images.  In the case of the MIPS 70 and 160
$\mu$m observations, these images represent the data from which the
photometry in the POPC is derived.  In the case of the IRAC and MIPS
24 $\mu$m data, photometry is derived from individual frames and the
results in the POPC are the median values with associated errors as
described above.

Low resolution (R $\sim$ 64--128) spectra obtained with the IRS are
presented in the spectral atlas comprised of extracted spectra from
7.4 - 33 $\mu$m for all stars and spectra from 5.2 - 33 $\mu$m for
3-30 Myr stars.  Again, because of on-orbit sensitivities, we chose to
drop the second order of the short low module (providing spectra from
5.2--7.4 $\mu$m) for the older sources in the sample rather than
decrease the number of stars in our program.  We also plan to deliver
an atlas of high resolution (R$\sim$600) IRS spectra comprised of data
from 9.9 to 37.2 $\mu$m for 35 stars chosen from amongst
our full sample.  Tables of emission line fluxes (or upper limits) are
provided for six features selected as most sensitive to remnant gas
based on the models of Gorti \& Hollenbach (2004).

\subsection{Calibration Products}

A primary product delivered for all sources in the survey 
are models of the stellar photosphere fit to 
short wavelength photometry and extended through the Spitzer
wavelengths.
The photospheric emission component is modeled by fitting
Kurucz atmospheres including convective overshoot to available
$BV$ Johnson, $vby$ Stromgren, $B_T V_T$
Tycho, $H_p$ Hipparcos, $RI$ Cousins, and $JHK_s$ 2MASS
photometry.
Predicted magnitudes were computed as described in
Cohen et al. (2003,and references therein)
using the combined system response of filter,
atmosphere (for ground-based observations), and detector.
The best-fit Kurucz model was computed in a least squares sense with
the
effective temperature and normalization constant (i.e. solid angle of
the 
source physically set by the distance and radius) as free
parameters, [Fe/H] fixed to solar metallicity, and surface gravity
fixed to
the value appropriate for the adopted stellar age and mass. Visual
extinction is fixed to $A_V = 0^m$ for stars with distances
less than 40~pc, assumed to be within the dust--free Local Bubble, but
a free parameter for stars at larger distances.  A file containing the
best--fit stellar spectral energy distribution (SED) 
is provided along with associated uncertainties in the fitted
parameters.   These fits are 
available from the Spitzer Legacy Science Archive
(http://ssc.spitzer.caltech.edu/legacy/all.html). 

Our secondary products include reported information concerning 
instrument calibration based on observations of FEPS
targets that are consistent with the expected level of stellar
photospheric emission.  We also investigate the consistency
between different instruments such as: 1) IRAC photometry at 5.4 
and 8.0 $\mu$m and the IRS spectra from 5.2--10 $\mu$m; 
2) MIPS 24 $\mu$m photometry and IRS spectra from 20--26 $\mu$m; 
and 3) low resolution and high resolution spectra from 
9.9--33 $\mu$m.  Details concerning these diagnostic comparisons
can be found in the FEPS Explanatory Supplements that accompany 
our data releases to the Spitzer Science Center. 

\section{Summary of Results to Date}

We briefly summarize the results from the FEPS program to date.
Silverstone et al. (2006) report a search for warm dust excesses 
surrounding 74 sun--like stars with ages 3--30 Myr.  Only five
objects show evidence for excess emission between 3.6--8 $\mu$m. 
All appear to be optically--thick disks and four shown signs of
active accretion from the disk onto the star.  This result suggests
that the transition time between optically--thick to optically--thin
inside of 1 AU is very short ($<$ 1 Myr).   Bouwman et al. (submitted) 
analyze the dust size and composition in the optically--thick accretion
disks in the FEPS sample from IRS high resolution observations. 
They report a correlation between the inferred grain size and
slope of the SED tracing disk structure.  They also analyze
the contribution of crystalline silicate emission to the 
observed spectra comparing the results to models for the production
of crystalline grains in the disk.  Hollenbach et al. 
(2005) report analysis of the IRS high resolution data for HD 105 
(30 Myr old) indicating that less than 0.1 M$_{JUPITER}$
of gas persists between 1-40 AU.  Extending this work, Pascucci 
et al. (2006) report similar results for a sample of 15 stars
spanning a range of age from 3--100 Myr.  It appears that gas--rich
disks capable of forming Jupiter--mass planets dissipate in less
than 10 Myr.   Additional observations planned will address whether
gas--rich disks persist beyond 3 Myr. 

Hines et al. (2006) report the discovery of an unusual warm 
debris disk around the 30 Myr old star HD 12039.  Assuming the 
excess is produced from dust dominated by large black--body grains, 
the emitting area is estimated to be between 4--6 AU from the 
star, comparable
to the location of our own asteroidal debris belt.  Stauffer
et al. (2005) analyze the frequency of 24 $\mu$m excess emission
among sun--like stars in the 100 Myr old Pleiades open cluster. 
They find a small fraction of stars exhibit excess emission
attributable to warm dust in the terrestrial planet zone.  Future
work will assess the fraction of warm dust excess as a function
of age throughout the FEPS sample. 

Initial discoveries of cool dust debris (Meyer et al. 2004; Kim 
et al. 2005) around FEPS targets suggest that: 1) there is large
dispersion of inferred dust masses in outer debris belts at any
one time; 2) there is a general trend of less dust at later times; 
and 3) most of these systems have large inner holes in their radial 
dust distributions.  Inner holes of order 30 AU in these disks with large 
dust mass surface density are probably maintained by mutual collisions of 
grains whose size are diminished down to the blow--out size whence they are
removed from the system due to radiation pressure (Dominik 
and Decin, 2003; Wyatt, 2005), although we cannot rule out
the presence of gas giant planets in most systems.  In an analysis
of a possible correlation between the presence of debris and
radial velocity planets, Moro--Martin et al. (submitted) report
no correlation in the FEPS database nor in the published suveys
of Bryden et al. (2006), as well as the detection of a debris
disk surrounding planet host star HD 38529.  Future 
work will focus on the fraction of objects with excess emission, 
the evolution in the mean dust mass as a function of age, and
the presence of extended debris disks around some stars (Hillenbrand
et al., in preparation), as well as 
connections between the presence of debris and 
metallicity of the central star. 

We would like to thank our colleagues at mission control at JPL, 
the Spitzer Science Center, as well as members of the IRAC, IRS, and MIPS 
instrument teams for their contributions to this work.   This work is based
in part on observations made with the Spitzer Space Telescope, which is operated
by the Jet Propulsion Laboratory, California Institute of Technology under
NASA contract 1407.  FEPS is pleased to acknowledge support through NASA
contracts 1224768, 1224634, and 1224566
administered through JPL.  MRM is also supported through 
membership in NASA's Astrobiology Institute. 
S.W. was supported by the German Research 
Foundation (DFG) though the Emmy Noether grant
WO 857/2--1.  EEM is supported by a Clay Fellowship from the 
Smithsonian Astrophysical Observatory.  The MPIA team is supported through the 
European Planet Formation Network funded by the EU.

\begin{deluxetable}{lccc}
\tablewidth{0pc}
\tablecaption{Field Stars}
\tablehead{\colhead{Name} & \colhead{RA(J2000)}  & \colhead{DEC(J2000)} & \colhead{Spectral Type}\\
&(J2000)&(J2000)& }
\startdata 
HD 224873 & 00:01:23.66 & +39:36:38.12 & K0\\
HD 377\tablenotemark{a} & 00:08:25.74 & +06:37:00.50 & G2V\\
HD 691 & 00:11:22.44 & +30:26:58.52 & K0V\\
HD 984 & 00:14:10.25 & -07:11:56.92 & F7V\\
HD 6434 & 01:04:40.15 & -39:29:17.61 & G2/3V\\
HD 6963 & 01:10:41.91 & +42:55:54.50 & G7V\\
HD 7661 & 01:16:24.19 & -12:05:49.33 & K0V\\
HIP 6276 & 01:20:32.27 & -11:28:03.74 & G0\\
HD 8941 & 01:28:24.36 & +17:04:45.20 & F8IV-V\\
HD 9472 & 01:33:19.03 & +23:58:32.19 & G0\\
HD 11850 & 01:56:47.27 & +23:03:04.09 & G5\\
HD 12039\tablenotemark{a} & 01:57:48.98 & -21:54:05.32 & G3/5V\\
HD 13382 & 02:11:23.15 & +21:22:38.39 & G5V\\
HD 13507 & 02:12:55.00 & +40:40:06.00 & G5V\\
HD 13531 & 02:13:13.35 & +40:30:27.34 & G7V\\
HD 13974 & 02:17:03.23 & +34:13:27.32 & G0V\\
HD 18940 & 03:03:28.65 & +23:03:41.19 & G0\\
HD 19019 & 03:03:50.82 & +06:07:59.82 & F8\\
HD 19668\tablenotemark{a} & 03:09:42.28 & -09:34:46.46 & G8/K0V\\
HD 21411 & 03:26:11.11 & -30:37:04.13 & G8V\\
HD 26990 & 04:16:16.50 & +07:09:34.15 & G0(V)\\
HD 27466 & 04:19:57.08 & -04:26:19.60 & G5V\\
HD 28495 & 04:33:54.23 & +64:37:59.40 & G0\\
HD 29231 & 04:34:38.49 & -35:39:29.06 & G8V\\
HD 31143 & 04:51:45.71 & -35:50:24.97 & K0V\\
HD 31392 & 04:54:04.21 & -35:24:16.28 & K0V\\
HD 32850 & 05:06:42.21 & +14:26:46.42 & G9V\\
HD 37572 & 05:36:56.86 & -47:57:52.87 & K0V\\
HD 37216 & 05:39:52.33 & +52:53:50.83 & G5\\
HD 37962 & 05:40:51.97 & -31:21:03.95 & G5V\\
HD 37006 & 05:46:11.89 & +78:15:22.61 & G0\\
HD 38529 & 05:46:34.92 & +01:10:05.31 & G8III/IV\\
HD 38949 & 05:48:20.06 & -24:27:50.04 & G1V\\
HD 40647 & 06:06:05.68 & +69:28:34.02 & G5\\
HD 43989 & 06:19:08.05 & -03:26:20.39 & G0V\\
HD 44594\tablenotemark{b} & 06:20:06.16 & -48:44:28.05 & G3V\\
HD 45270 & 06:22:30.97 & -60:13:07.14 & G1V\\
HD 61005\tablenotemark{a} & 07:35:47.47 & -32:12:14.11 & G3/5V\\
HD 60737 & 07:38:16.44 & +47:44:55.34 & G0\\
HD 61994 & 07:47:30.61 & +70:12:23.97 & G6V\\
HD 64324 & 07:54:48.47 & +34:37:11.42 & G0\\
HD 66751 & 08:10:20.51 & +69:43:30.21 & F8V\\
HD 69076 & 08:15:07.73 & -06:55:08.23 & K0V\\
HD 70516 & 08:24:15.66 & +44:56:58.92 & G0\\
HD 71974 & 08:31:35.05 & +34:57:58.44 & G5\\
HD 72687 & 08:33:15.39 & -29:57:23.66 & G5V\\
HD 73668 & 08:39:43.81 & +05:45:51.59 & G1V\\
HIP 42491 & 08:39:44.69 & +05:46:14.00 & G5\\
HD 75302 & 08:49:12.53 & +03:29:05.25 & G5V\\
HD 75393 & 08:49:15.35 & -15:33:53.12 & F7V\\
HD 76218 & 08:55:55.68 & +36:11:46.40 & G9-V\\
HD 77407 & 09:03:27.08 & +37:50:27.72 & G0(V)\\
HD 80606 & 09:22:37.56 & +50:36:13.43 & G5\\
HD 85301 & 09:52:16.77 & +49:11:26.84 & G5\\
HD 88201 & 10:09:31.78 & -32:50:47.95 & G0V\\
HD 88742 & 10:13:24.72 & -33:01:54.22 & G0V\\
HD 90712 & 10:27:47.79 & -34:23:58.14 & G2/3V\\
HD 90905 & 10:29:42.23 & +01:29:27.82 & G1V\\
HD 91782 & 10:36:47.84 & +47:43:12.42 & G0\\
HD 91962 & 10:37:00.02 & -08:50:23.63 & G1V\\
HD 92788 & 10:42:48.54 & -02:11:01.38 & G6V\\
HD 92855 & 10:44:00.62 & +46:12:23.86 & F9V\\
HD 95188 & 10:59:48.28 & +25:17:23.65 & G8V\\
HD 98553 & 11:20:11.60 & -19:34:40.54 & G2/3V\\
HD 100167 & 11:31:53.92 & +41:26:21.65 & F8\\
HD 101472 & 11:40:36.59 & -08:24:20.32 & F7V\\
HD 101959 & 11:43:56.62 & -29:44:51.80 & G0V\\
HD 102071 & 11:44:39.32 & -29:53:05.46 & K0V\\
HD 103432 & 11:54:32.07 & +19:24:40.44 & G6V\\
HD 104576 & 12:02:39.46 & -10:42:49.16 & G3V\\
HD 104860 & 12:04:33.71 & +66:20:11.58 & F8\\
HD 105631 & 12:09:37.26 & +40:15:07.62 & G9V\\
HD 106156 & 12:12:57.52 & +10:02:15.62 & G8V\\
HD 106252 & 12:13:29.49 & +10:02:29.96 & G0\\
HD 107146 & 12:19:06.49 & +16:32:53.91 & G2V\\
HD 108799 & 12:30:04.77 & -13:23:35.14 & G1/2V\\
HD 108944 & 12:31:00.74 & +31:25:25.84 & F9V\\
HD 112196 & 12:54:40.02 & +22:06:28.65 & F8V\\
HD 115043 & 13:13:37.01 & +56:42:29.82 & G1V\\
HD 121320 & 13:54:28.20 & +20:38:30.46 & G5V\\
HD 121504 & 13:57:17.23 & -56:02:24.27 & G2V\\
HD 122652 & 14:02:31.63 & +31:39:39.09 & F8\\
HD 129333 & 14:39:00.25 & +64:17:29.94 & G5V\\
HD 132173 & 14:58:30.51 & -28:42:34.15 & G0V\\
HD 133295 & 15:04:33.08 & -28:18:00.65 & G0/1V\\
HD 136923 & 15:22:46.84 & +18:55:08.31 & G9V\\
HD 138004 & 15:27:40.36 & +42:52:52.82 & G2III\\
HD 139813 & 15:29:23.61 & +80:27:01.08 & G5\\
HD 141937 & 15:52:17.55 & -18:26:09.80 & G2/3V\\
HD 142229 & 15:53:20.02 & +04:15:11.51 & G5V\\
HD 145229 & 16:09:26.63 & +11:34:28.25 & G0\\
HD 150706 & 16:31:17.63 & +79:47:23.15 & G3(V)\\
HD 150554 & 16:40:56.45 & +21:56:53.24 & F8\\
HD 151798 & 16:50:05.17 & -12:23:14.88 & G3V\\
HD 152555 & 16:54:08.15 & -04:20:24.89 & F8/G0V\\
HD 153458 & 17:00:01.66 & -07:31:53.93 & G5V\\
HD 154417 & 17:05:16.83 & +00:42:09.18 & F9V\\
HD 157664\tablenotemark{b} & 17:18:58.47 & +68:52:40.61 & G0\\
HD 159222 & 17:32:00.99 & +34:16:15.97 & G1V\\
HD 161897 & 17:41:06.70 & +72:25:13.41 & K0\\
HD 167389 & 18:13:07.22 & +41:28:31.33 & F8(V)\\
HD 170778 & 18:29:03.94 & +43:56:21.54 & G5\\
HD 172649 & 18:39:42.11 & +37:59:35.22 & F5\\
HD 179949 & 19:15:33.23 & -24:10:45.61 & F8V\\
HD 183216 & 19:29:40.57 & -30:47:52.36 & G2V\\
HD 187897 & 19:52:09.38 & +07:27:36.10 & G5\\
HD 190228 & 20:03:00.77 & +28:18:24.46 & G5IV\\
HD 193017 & 20:18:10.00 & -04:43:43.23 & F6V\\
HD 195034 & 20:28:11.81 & +22:07:44.34 & G5\\
HD 199019 & 20:49:29.30 & +71:46:29.29 & G5\\
HD 199598 & 20:57:39.68 & +26:24:18.40 & G0V\\
HD 200746 & 21:05:07.95 & +07:56:43.59 & G5\\
HD 201219 & 21:07:56.53 & +07:25:58.47 & G5\\
HD 202108 & 21:12:57.63 & +30:48:34.25 & G3V\\
HD 201989 & 21:14:01.80 & -29:39:48.85 & G3/5V\\
HD 203030 & 21:18:58.22 & +26:13:50.05 & G8V\\
HD 204277 & 21:27:06.61 & +16:07:26.85 & F8V\\
HD 205905 & 21:39:10.14 & -27:18:23.59 & G2V\\
HD 206374 & 21:41:06.19 & +26:45:02.25 & G6.5V\\
HD 209393 & 22:02:05.38 & +44:20:35.47 & G5\\
HD 209779 & 22:06:05.32 & -05:21:29.15 & G2V\\
HD 212291 & 22:23:09.17 & +09:27:39.95 & G5\\
HD 216275 & 22:50:46.34 & +52:03:41.21 & G0\\
HD 217343 & 23:00:19.29 & -26:09:13.48 & G3V
\enddata
\tablenotetext{a}{IRS high resolution spectra were also obained
for these stars.}
\tablenotetext{b}{These stars also also Spitzer Space Telescope calibration targets.}
\end{deluxetable}

\begin{deluxetable}{lcccc}
\tablewidth{0pc}
\tablecaption{Open Cluster Stars}
\tablehead{\colhead{Name} & \colhead{RA(J2000)}  & \colhead{DEC(J2000)} & \colhead{Spectral Type} & \colhead{Open Cluster}}
\startdata
vB 1 & 03:17:26.39 & +07:39:20.90 & F8 & Hyades\\
HE 350 & 03:17:36.93 & +48:50:08.50 &  -  & Alpha Per\\
HE 373 & 03:18:27.39 & +47:21:15.42 &  -  & Alpha Per\\
HE 389 & 03:18:50.31 & +49:43:52.19 &  -  & Alpha Per\\
AP 93 & 03:19:02.76 & +48:10:59.61 &  -  & Alpha Per\\
HE 622 & 03:24:49.71 & +48:52:18.33 &  -  & Alpha Per\\
HE 696 & 03:26:19.36 & +49:13:32.54 &  -  & Alpha Per\\
HE 699 & 03:26:22.22 & +49:25:37.52 &  -  & Alpha Per\\
HE 750 & 03:27:37.79 & +48:59:28.78 & F5 & Alpha Per\\
HE 767 & 03:27:55.02 & +49:45:37.16 &  -  & Alpha Per\\
HE 848 & 03:29:26.24 & +48:12:11.74 & F9V & Alpha Per\\
HE 935 & 03:31:28.99 & +48:59:28.37 & F9.5V & Alpha Per\\
HE 1101 & 03:35:08.75 & +49:44:39.59 &  -  & Alpha Per\\
HE 1234 & 03:39:02.91 & +51:36:37.11 &  -  & Alpha Per\\
HII 120 & 03:43:31.95 & +23:40:26.61 &  -  & Pleiades\\
HII 152 & 03:43:37.73 & +23:32:09.59 & G5V & Pleiades\\
HII 174 & 03:43:48.33 & +25:00:15.83 &  -  & Pleiades\\
HII 173 & 03:43:48.41 & +25:11:24.19 &  -  & Pleiades\\
HII 250 & 03:44:04.24 & +24:59:23.40 &  -  & Pleiades\\
HII 314 & 03:44:20.09 & +24:47:46.16 &  -  & Pleiades \\
HII 514 & 03:45:04.01 & +25:15:28.23 &  -  & Pleiades\\
HII 1015 & 03:46:27.35 & +25:08:07.97 &  -  & Pleiades\\
HII 1101 & 03:46:38.78 & +24:57:34.61 & G0V & Pleiades\\
HII 1182 & 03:46:47.06 & +22:54:52.48 &  -  & Pleiades\\
HII 1200 & 03:46:50.54 & +23:14:21.06 &  -  & Pleiades\\
HII 1776 & 03:48:17.70 & +25:02:52.29 &  -  & Pleiades\\
HII 2147 & 03:49:06.11 & +23:46:52.49 & G7IV & Pleiades \\
HII 2278 & 03:49:25.70 & +24:56:15.43 &  -  & Pleiades\\
HII 2506 & 03:49:56.49 & +23:13:07.01 &  -  & Pleiades\\
HII 2644 & 03:50:20.90 & +24:28:00.22 &  -  & Pleiades\\
HII 2786 & 03:50:40.08 & +23:55:58.94 &  -  & Pleiades\\
HII 2881 & 03:50:54.32 & +23:50:05.52 & K2 & Pleiades\\
HII 3097 & 03:51:40.44 & +24:58:59.41 &  -  & Pleiades\\
HII 3179 & 03:51:56.86 & +23:54:06.98 &  -  & Pleiades\\
vB 39 & 04:22:44.74 & +16:47:27.56 & G4V & Hyades\\
vB 49 & 04:24:12.78 & +16:22:44.22 & G0V & Hyades\\
vB 52 & 04:24:28.33 & +16:53:10.32 & G2V & Hyades\\
vB 176 & 04:25:47.56 & +18:01:02.20 & K2V & Hyades\\
vB 63 & 04:26:24.61 & +16:51:11.84 & G1V & Hyades\\
vB 64 & 04:26:40.11 & +16:44:48.78 & G2+ & Hyades\\
vB 66 & 04:27:46.07 & +11:44:11.07 & F8 & Hyades\\
vB 73 & 04:28:48.29 & +17:17:07.84 & G2V & Hyades\\
vB 79 & 04:29:31.61 & +17:53:35.46 & K0V & Hyades\\
vB 180 & 04:29:57.73 & +16:40:22.23 & K1V & Hyades\\
vB 88 & 04:31:29.35 & +13:54:12.55 & F9V & Hyades\\
vB 91 & 04:32:50.12 & +16:00:20.96 &  -  & Hyades\\
vB 92 & 04:32:59.45 & +15:49:08.37 &  -  & Hyades\\
vB 93 & 04:33:37.97 & +16:45:44.96 &  -  & Hyades\\
vB 96 & 04:33:58.54 & +15:09:49.04 & G5 & Hyades\\
vB 183 & 04:34:32.18 & +15:49:39.23 &  -  & Hyades\\
vB 97 & 04:34:35.31 & +15:30:16.56 & F8:V & Hyades\\
vB 99 & 04:36:05.27 & +15:41:02.60 &  -  & Hyades\\
vB 106 & 04:38:57.31 & +14:06:20.16 & G5 & Hyades\\
vB 142 & 04:46:30.38 & +15:28:19.38 & G5 & Hyades\\
vB 143 & 04:51:23.22 & +15:26:00.45 & F8 & Hyades\\
R3 & 10:29:32.75 & -63:49:15.68 &  -  & IC2602\\
R45 & 10:40:00.03 & -63:15:11.04 &  -  & IC2602\\
W79 & 10:42:07.07 & -64:46:07.85 &  -  & IC2602\\
B102 & 10:42:41.52 & -64:21:04.37 &  -  & IC2602\\
R83 & 10:46:14.83 & -64:02:58.05 &  -  & IC2602
\enddata
\end{deluxetable}

\begin{deluxetable}{lccc}
\tablewidth{0pc}
\tablecaption{Young Stars}
\tablehead{\colhead{Name} & \colhead{RA(J2000)}  & \colhead{DEC(J2000)} & \colhead{Spectral Type} }
\startdata
HD 105\tablenotemark{a} & 00:05:52.56 & -41:45:10.98 & G0V\\
QT And & 00:41:17.32 & +34:25:16.77 & G\\
RE J0137+18A & 01:37:39.41 & +18:35:33.16 & K3Ve\\
HD 15526 & 02:29:35.03 & -12:24:08.56 & G5/6V\\
1RXS J025216.9+361658 & 02:52:17.59 & +36:16:48.14 & K2IV\\
2RE J0255+474 & 02:55:43.60 & +47:46:47.58 & K5Ve\\
1RXS J025751.8+115759 & 02:57:51.68 & +11:58:05.83 & G7V\\
RX J0258.4+2947 & 02:58:28.77 & +29:47:53.80 & K0IV\\
1RXS J030759.1+302032 & 03:07:59.20 & +30:20:26.05 & G5IV\\
1E 0307.4+1424 & 03:10:12.55 & +14:36:02.90 & G6V\\
1RXS J031644.0+192259 & 03:16:43.89 & +19:23:04.11 & G2V\\
1RXS J031907.4+393418 & 03:19:07.61 & +39:34:10.50 & K0V\\
1E 0324.1-2012 & 03:26:22.05 & -20:01:48.81 & G4V\\
RX J0329.1+0118 & 03:29:08.06 & +01:18:05.66 & G0(IV)\\
RX J0331.1+0713 & 03:31:08.38 & +07:13:24.78 & K4(V)/E\\
HD 22179 & 03:35:29.91 & +31:13:37.45 & G5IV\\
1RXS J034423.3+281224 & 03:44:24.25 & +28:12:23.07 & G7V\\
1RXS J035028.0+163121 & 03:50:28.40 & +16:31:15.19 & G5IV\\
RX J0354.4+0535 & 03:54:21.31 & +05:35:40.77 & G2(V)\\
RX J0357.3+1258 & 03:57:21.39 & +12:58:16.83 & G0\\
HD 25300 & 03:59:36.73 & -39:53:14.85 & K0\\
HD 285281 & 04:00:31.07 & +19:35:20.70 & K1\\
HD 285372 & 04:03:24.95 & +17:24:26.12 & K3(V)\\
HD 284135\tablenotemark{a} & 04:05:40.58 & +22:48:12.14 & G3(V)\\
HD 281691 & 04:09:09.74 & +29:01:30.55 & K1(V)\\
HD 26182 & 04:10:04.69 & +36:39:12.14 & G0V\\
HD 284266 & 04:15:22.92 & +20:44:16.93 & K0(V)\\
HD 285751 & 04:23:41.33 & +15:37:54.87 & K2(V)\\
HD 279788 & 04:26:37.40 & +38:45:02.37 & G5V\\
HD 285840 & 04:32:42.43 & +18:55:10.25 & K1(V)\\
1RXS J043243.2-152003 & 04:32:43.51 & -15:20:11.39 & G4V\\
RX J0434.3+0226 & 04:34:19.54 & +02:26:26.10 & K4e\\
HD 282346 & 04:39:31.00 & +34:07:44.43 & G8V\\
RX J0442.5+0906 & 04:42:32.09 & +09:06:00.86 & G5(V)\\
HD 31281 & 04:55:09.62 & +18:26:30.84 & G1(V)\\
HD 286179 & 04:57:00.65 & +15:17:53.09 & G3(V)\\
HD 31950 & 05:00:24.31 & +15:05:25.28 &  - \\
HD 286264 & 05:00:49.28 & +15:27:00.68 & K2IV\\
1RXS J051111.1+281353 & 05:11:10.53 & +28:13:50.38 & K0V\\
1RXS J053650.0+133756 & 05:36:50.06 & +13:37:56.22 & K0V\\
HD 245567 & 05:37:18.44 & +13:34:52.52 & G0V\\
SAO 150676 & 05:40:20.74 & -19:40:10.85 & G2V\\
AO Men\tablenotemark{a} & 06:18:28.24 & -72:02:41.56 & K3.5\\
HD 47875 & 06:34:41.04 & -69:53:06.35 & G3V\\
RE J0723+20 & 07:23:43.58 & +20:24:58.64 & K3(V)\\
HD 70573 & 08:22:49.95 & +01:51:33.58 & G1/2V\\
RX J0849.2-7735 & 08:49:11.11 & -77:35:58.53 & K1(V)\\
RX J0850.1-7554 & 08:50:05.41 & -75:54:38.11 & G5\\
RX J0853.1-8244 & 08:53:05.29 & -82:43:59.71 & K0(V)\\
RX J0917.2-7744 & 09:17:10.33 & -77:44:01.99 & G2\\
HD 86356 & 09:51:50.70 & -79:01:37.73 & G6/K0\\
SAO 178272 & 09:59:08.42 & -22:39:34.57 & K2V\\
MML 1 & 10:57:49.37 & -69:13:59.99 & K1+IV\\
RX J1111.7-7620\tablenotemark{a} & 11:11:46.32 & -76:20:09.21 & K1\\
RX J1140.3-8321 & 11:40:16.59 & -83:21:00.38 & K2\\
BPM 87617 & 11:47:45.73 & +12:54:03.31 & K5Ve\\
HD 104467 & 12:01:39.15 & -78:59:16.85 & G5III/IV\\
RX J1203.7-8129 & 12:03:24.70 & -81:29:55.28 & K1\\
HIP 59154 & 12:07:51.19 & -75:55:15.97 & K2\\
RX J1209.8-7344 & 12:09:42.82 & -73:44:41.41 & G9\\
MML 8\tablenotemark{a} & 12:12:35.77 & -55:20:27.31 & K0+IV\\
MML 9 & 12:14:34.10 & -51:10:12.47 & G9IV\\
HD 106772 & 12:17:26.94 & -80:35:06.90 & G2III/IV\\
RX J1220.6-7539 & 12:20:34.38 & -75:39:28.65 & K2\\
HD 107441 & 12:21:16.48 & -53:17:45.06 & G1.5IV\\
MML 17\tablenotemark{a} & 12:22:33.23 & -53:33:48.95 & G0IV\\
MML 18 & 12:23:40.13 & -56:16:32.57 & K0+IV\\
RX J1225.3-7857 & 12:25:13.40 & -78:57:34.71 & G5\\
HD 111170 & 12:47:51.86 & -51:26:38.29 & G8/K0V\\
MML 26 & 12:48:48.19 & -56:35:37.90 & G5IV\\
MML 28\tablenotemark{a} & 13:01:50.70 & -53:04:58.11 & K2-IV\\
MML 32 & 13:17:56.94 & -53:17:56.21 & G1IV\\
HD 116099 & 13:22:04.47 & -45:03:23.19 & G0/3\\
PDS 66\tablenotemark{a} & 13:22:07.53 & -69:38:12.18 & K1IVe\\
HD 117524 & 13:31:53.61 & -51:13:33.05 & G2.5IV\\
MML 36\tablenotemark{a} & 13:37:57.30 & -41:34:41.98 & K0IV\\
HD 119269 & 13:43:28.54 & -54:36:43.44 & G3/5V\\
MML 38 & 13:47:50.55 & -49:02:05.61 & G8IVe\\
HD 120812 & 13:52:47.80 & -46:44:09.24 & F8/G0V\\
MML 40 & 14:02:20.72 & -41:44:50.93 & G9IV\\
MML 43 & 14:27:05.56 & -47:14:21.73 & G7IV\\
HD 126670 & 14:28:09.30 & -44:14:17.54 & G6/8III/IV\\
HD 128242 & 14:37:04.22 & -41:45:02.91 & G3V\\
RX J1450.4-3507 & 14:50:25.82 & -35:06:48.66 & K1(IV)\\
MML 51 & 14:52:41.98 & -41:41:55.24 & K1IVe\\
RX J1457.3-3613 & 14:57:19.62 & -36:12:27.44 & G6IV\\
RX J1458.6-3541 & 14:58:37.69 & -35:40:30.27 & K3(IV)\\
RX J1500.8-4331 & 15:00:51.89 & -43:31:21.23 & K1(IV)\\
MML 57 & 15:01:58.82 & -47:55:46.46 & G1.5IV\\
RX J1507.2-3505 & 15:07:14.81 & -35:04:59.55 & K0\\
HD 135363 & 15:07:56.31 & +76:12:02.66 & G5(V)\\
HD 133938 & 15:08:38.50 & -44:00:51.99 & G6/8III/IV\\
RX J1518.4-3738 & 15:18:26.92 & -37:38:02.14 & K1\\
RX J1531.3-3329 & 15:31:21.93 & -33:29:39.46 & K0\\
HIP 76477 & 15:37:11.30 & -40:15:56.70 & G9\\
V343 Nor\tablenotemark{a} & 15:38:57.57 & -57:42:27.30 & K0V\\
HD 139498 & 15:39:24.40 & -27:10:21.87 & G8(V)\\
RX J1541.1-2656 & 15:41:06.79 & -26:56:26.33 & G7\\
RX J1544.0-3311 & 15:44:03.76 & -33:11:11.09 & K1\\
HD 140374 & 15:44:21.06 & -33:18:54.97 & G8V\\
RX J1545.9-4222 & 15:45:52.25 & -42:22:16.41 & K1\\
HD 141521 & 15:51:13.74 & -42:18:51.36 & G8V\\
HD 141943\tablenotemark{a} & 15:53:27.29 & -42:16:00.81 & G0/2V\\
HD 142361\tablenotemark{a} & 15:54:59.86 & -23:47:18.26 & G3V\\
$[$PZ99$]$ J155847.8-175800 & 15:58:47.73 & -17:57:59.58 & K3\\
RX J1600.6-2159\tablenotemark{a} & 16:00:40.57 & -22:00:32.24 & G9\\
HD 143358 & 16:01:07.93 & -32:54:52.65 & G1/2V\\
ScoPMS 21 & 16:01:25.63 & -22:40:40.38 & K1IV\\
ScoPMS 27 & 16:04:47.76 & -19:30:23.12 & K2IV\\
$[$PZ99$]$ J160814.7-190833 & 16:08:14.74 & -19:08:32.77 & K2\\
ScoPMS 52\tablenotemark{a} & 16:12:40.51 & -18:59:28.31 & K0IV\\
$[$PZ99$]$ J161318.6-221248 & 16:13:18.59 & -22:12:48.96 & G9\\
$[$PZ99$]$ J161329.3-231106 & 16:13:29.29 & -23:11:07.56 & K1\\
$[$PZ99$]$ J161402.1-230101 & 16:14:02.12 & -23:01:02.18 & G4\\
$[$PZ99$]$ J161411.0-230536\tablenotemark{a} & 16:14:11.08 & -23:05:36.26 & K0\\
$[$PZ99$]$ J161459.2-275023\tablenotemark{a} & 16:14:59.18 & -27:50:23.06 & G5\\
$[$PZ99$]$ J161618.0-233947 & 16:16:17.95 & -23:39:47.70 & G7\\
HD 146516 & 16:17:31.39 & -23:03:36.02 & G0IV\\
ScoPMS 214\tablenotemark{a} & 16:29:48.70 & -21:52:11.91 & K0IV\\
RX J1839.0-3726 & 18:39:05.29 & -37:26:21.78 & K1\\
RX J1841.8-3525 & 18:41:48.56 & -35:25:43.71 & G7\\
RX J1842.9-3532\tablenotemark{a} & 18:42:57.98 & -35:32:42.73 & K2\\
RX J1844.3-3541 & 18:44:21.92 & -35:41:43.53 & K5\\
RX J1852.3-3700\tablenotemark{a} & 18:52:17.30 & -37:00:11.93 & K3\\
HD 174656 & 18:53:05.99 & -36:10:22.91 & G6IV\\
RX J1917.4-3756\tablenotemark{a} & 19:17:23.83 & -37:56:50.52 & K2\\
HD 199143 & 20:55:47.68 & -17:06:51.02 & F8V\\
V383 Lac & 22:20:07.03 & +49:30:11.67 & K0IV/V\\
RX J2313.0+2345 & 23:13:01.24 & +23:45:29.64 & F8\\
HD 219498 & 23:16:05.02 & +22:10:34.98 & G5
\enddata
\tablenotetext{a}{IRS high resolution spectra were also obained
for these stars.}
\end{deluxetable}

\begin{deluxetable}{lccc}
\tablewidth{0pc}
\tablecaption{Pre--selected IRS High Resolution Targets}
\tablehead{\colhead{Name} & \colhead{RA(J2000)}  & \colhead{DEC(J2000)} & \colhead{Spectral Type} }
\startdata
HD 8907 & 01:28:34.35 & +42:16:03.70 & F8\\
HD 17925 & 02:52:32.14 & -12:46:11.18 & K1V\\
HD 25457 & 04:02:36.76 & -00:16:08.17 & F7V\\
HD 35850 & 05:27:04.77 & -11:54:03.38 & F7/8V\\
HD 37484 & 05:37:39.63 & -28:37:34.65 & F3V\\
HD 38207\tablenotemark{a} & 05:43:20.95 & -20:11:21.41 & F2V\\
HD 41700 & 06:04:28.44 & -45:02:11.71 & F8/G0V\\
HD 72905\tablenotemark{a} & 08:39:11.62 & +65:01:15.14 & G1.5VB\\
HD 134319 & 15:05:49.90 & +64:02:50.00 & G5(V)\\
HD 143006 & 15:58:36.92 & -22:57:15.35 & G6/8\\
HD 191089\tablenotemark{a} & 20:09:05.22 & -26:13:26.63 & F5V\\
HD 202917 & 21:20:49.95 & -53:02:03.05 & G5V\\
HD 209253 & 22:02:32.97 & -32:08:01.60 & F6/7V\\
HD 216803 & 22:56:24.07 & -31:33:56.12 & K4VP
\enddata
\tablenotetext{a}{IRS high resolution spectra were not
obtained for these sources due to program constraints.
}
\end{deluxetable}
\end{document}